\input epsf.tex
\documentclass[12pt]{article}
\usepackage{graphicx}
\usepackage{amssymb,epsf}
\csname @addtoreset\endcsname{equation}{section}

\def\bseq{\begin{subequation}}  
\def\eseq{\end{subequation}}
\def\Bar#1{\overline{#1}}                       

\def\Tilde#1{\widetilde{#1}}                    

\newcommand{\beq}{\begin{equation}}
\newcommand{\eeq}{\end{equation}}
\newcommand{\bea}{\begin{eqnarray}}
\newcommand{\eea}{\end{eqnarray}}
\newcommand{\ena}{\end{eqnarray}}

\newcommand {\non}{\nonumber}

\renewcommand{\a}{\alpha}
\renewcommand{\b}{\beta}

\renewcommand{\d}{\delta}
\renewcommand{\th}{\theta}

\newcommand{\pa}{\partial}
\newcommand{\g}{\gamma}
\newcommand{\G}{\Gamma}

\newcommand{\e}{\epsilon}

\renewcommand{\L}{\Lambda}

\newcommand{\Db}{\Bar{D}}
\newcommand{\Wb}{\Bar{W}}

\newcommand{\phib}{\bar{\phi}}

\newcommand{\thb}{\bar{\theta}}
\newcommand{\Phib}{\bar{\Phi}}

\newcommand{\Lb}{\bar{\Lambda}}
\newcommand{\Gbar}{\Bar{\Gamma}}
\newcommand{\Phibold}{\bold{\Phi}}
\newcommand{\Phibbold}{\bold{\bar{\Phi}}}
\newcommand{\barh}{\bar{h}}

\newcommand{\Nc}{{\cal{N}}}
\newcommand{\ad}{{\dot{\alpha}}}
\newcommand{\bd}{{\dot{\beta}}}
\newcommand{\gd}{{\dot{\gamma}}}

\newcommand{\Del}{\nabla}
\newcommand{\Delb}{\overline{\nabla}}

\newcommand{\Dela}{\nabla_{\alpha}}

\newcommand{\boldnabla}{  \nabla \hspace{-0.12in}{\nabla}}
\newcommand{\Wbar}{\overline{W}}
\newcommand{\Tr}{{\rm Tr}}

\newcommand{\intsup}{\int\!\! d^4xd^4\theta ~}     
\newcommand{\intch}{\int\!\! d^4xd^2\theta ~}      
\newcommand{\intach}{\int\!\! d^4xd^2\thb ~} 

\newcommand{\Fab}{\mathcal{F}^{\alpha\beta}}
\newcommand{\mathN}{\mathcal{N}}

\begin{document}
\begin{titlepage}
\begin{flushright}
CPHT-RR002.0109\\
LPT-Orsay 09/03
\end{flushright}
\vspace{.3cm}
\begin{center}
{\Large \bf A renormalizable $N=1/2$ SYM theory \\
with interacting matter}
\vfill

{\large \bf Silvia Penati$^1$,~Alberto Romagnoni$^2$, 
Massimo Siani$^1$}\\

\vspace{0.5cm}

{\small 
$^1$ Dipartimento di Fisica, Universit\`a di Milano--Bicocca and\\
INFN, Sezione di Milano--Bicocca, Piazza della
Scienza 3, I-20126 Milano, Italy\\

\vspace{0.1cm}
$^2$ Laboratoire de Physique Th«eorique, Univ. Paris-Sud and CNRS, F-91405 Orsay, and \\
CPhT, Ecole Polytechnique, CNRS, 91128 Palaiseau Cedex, France }\\

\end{center}
\vfill
\begin{center}
{\bf Abstract}
\end{center}
{We consider nonanticommutative SYM theories with chiral matter in the adjoint 
representation of the $SU({\cal N}) \otimes U(1)$ gauge group. 
In a superspace setup and manifest background covariant approach we investigate the one--loop
renormalization of the theory when a cubic superpotential is present. The structure of the divergent
terms reveals that the theory simply obtained from the ordinary one by trading products for 
star products is not renormalizable. Moreover, because of the different renormalization 
undergone by the abelian field compared to the non-abelian ones, the superpotential seems to be 
incompatible with the requests of renormalizability, gauge and $N=1/2$ invariance. However, 
by a suitable modification of the quadratic action for the $U(1)$ (anti)chiral superfields and the
addition of extra couplings,         
we find an action which is one-loop renormalizable and manifestly $N=1/2$ supersymmetric and supergauge 
invariant. We conclude that interacting matter can be safely introduced in NAC gauge theories, 
in contrast with previous results.}

\vspace{2mm} \vfill \hrule width 3.cm
\begin{flushleft}
e-mail: silvia.penati@mib.infn.it\\
e-mail: alberto.romagnoni@cpht.polytechnique.fr\\
e-mail: massimo.siani@mib.infn.it
\end{flushleft}
\end{titlepage}

\section{Introduction}

Non(anti)commutative field theories emerge naturally as low energy limits of strings  
in a background where a constant Neveu--Schwarz two form and/or a Ramond-Ramond two--form are turned on
\cite{SW, DV, BGV, seiberg}. In the supersymmetric case, the appearance of the RR flux ${\cal F}_{\a\b}$ 
modifies the superspace geometry through the appearance of a nontrivial anticommutator 
$\{\th_\a , \th_\b \} = {\cal F}_{\a\b}$ \cite{FL, KPT, FLM, BGV, seiberg, BFPL}. The effect on field theories defined
in nonanticommutative (NAC) superspace is that the multiplication among superfields is no longer 
commutative but described by the so-called $\ast$--product. As a result, supersymmetry is in general 
partially broken from $N=1$ to $N=1/2$. For extended supersymmetries suitable deformations can be realized
which break less supersymmetry \cite{ferraraetal}.   

In the recent past quantum properties of NAC theories have been investigated. In
particular, renormalizability is one of the more interesting issues since the partial 
breaking of supersymmetry could affect the ordinary boson-fermion cancellation leading to a worsening of the
UV behavior of the theory. For the deformed WZ model loop calculations have been performed both
in superspace \cite{TY,BFR2, GPR1} and in components \cite{JJP}. They reveal that renormalizability is
lost already at one loop but it can be restored by adding new couplings in the classical action depending
on the deformation parameter. This modification is then sufficient to make the theory renormalizable at all
orders \cite{BFR}. The same analysis has been carried on also for deformed $SU(\Nc) \otimes U(1)$ gauge theories 
in interaction with massive quantum chiral matter in the adjoint representation. In \cite{GPR2}
we have found a general action for the pure gauge sector which is $N=1/2$ supergauge invariant 
and one-loop renormalizable. It differs from the one obtained from the ordinary action where we trade 
products for $\ast$--products by the addition of new couplings depending on the deformation parameter. They
span the spectrum of all possible couplings allowed by supergauge invariance. Our results are confirmed
by a similar analysis done in components \cite{JJW}. General arguments in support of renormalizability 
for  $N=1/2$ gauge theories coupled to non--interacting matter can be found in \cite{LR}. 

It is important to stress that in all theories 
investigated so far UV divergences are always logarithmic. This suggests that under NAC deformations 
supersymmetry is in general softly broken. 

The previous results for super-Yang-Mills theories concern primarily the gauge sector. However, 
for a complete proof of the
quantum consistency of the theories one should analyze also the matter sector. A preliminary discussion
on theories with non-interacting massive matter can be found in \cite{GPR2}, whereas 
a systematic attempt in this direction has been carried on recently in \cite{JJW3}. 

Working in components in the WZ gauge, the authors of \cite{JJW3} have investigated the structure
of one--loop divergences in all sectors of the theory. When massive matter is present in the fundamental 
and/or in the adjoint representation of $SU(\Nc) \otimes U(1)$, both the gauge and matter sectors 
can be made finite by a suitable 
generalization of the classical action which contains new deformation-dependent couplings in addition 
to the ones obtained by generalizing products to $\ast$-products. 

For chiral matter in the adjoint representation one can also add a superpotential term. 
In \cite{JJW3}
a deformed SYM theory with cubic superpotential has been investigated. The authors have
found that the cancellation of one-loop divergences in the matter sector requires a modification of the 
classical superpotential which breaks supersymmetry completely. This is due to the following mechanism:
According to  the non-renormalization theorem, the renormalization of the chiral coupling is induced 
by the renormalization of the (anti)chiral superfields. On the other hand, abelian and non-abelian fields 
renormalize differently, so that in a $SU({\cal N}) \otimes U(1)$ theory a generalization of the 
superpotential which assigns different couplings to the abelian and non-abelian sectors is necessary 
in order to render the theory renormalizable.  
While in the ordinary case this is consistent with supersymmetry, in the NAC case one can easily 
realize that the generalized superpotential is no longer supersymmetric.  
Therefore, it seems that in NAC SYM theories interacting chiral matter can be consistently added at quantum level 
only at the price to give up supersymmetry completely. We will call it the ``superpotential problem''.

In particular, it follows that $N=4$ SYM does not seem to possess a renormalizable $N=1/2$ deformation.
On the other hand, string theory would provide a natural interpretation of this theory as 
the low energy dynamics of a set of D3--branes in a constant graviphoton background. Therefore, 
the absence of $N=1/2$ generalizations of $N=4$ SYM should be understood also from a string theory 
point of view. 

As a first step it is then important to investigate whether the negative results of Ref. \cite{JJW3} 
find definitive confirmation or they can be overpassed.  
To this end, in a superspace setup we reconsider the problem of quantizing NAC SYM theories with a
cubic superpotential. We start from the natural $\ast$--generalization of the ordinary superspace 
action for $N=1$ SYM with cubic superpotential for a single (anti)chiral field. First of all, 
we rephrase the conclusions of \cite{JJW3} in superspace language by arguing that the request for 
the theory to be renormalizable and supergauge invariant would force the appearance of  terms in 
the action which would be manifestly non--supersymmetric.

Successively, we prove that a suitable generalization of the action can be found which solves the 
superpotential problem. 
It is obtained by assigning a different coupling constant to the quadratic term for the abelian 
matter superfields. The modification is done in a manifestly $N=1/2$ supersymmetric and supergauge 
invariant way and has a double effect:
On one side, the kinetic terms for the abelian and non--abelian superfields appear with 
a different normalization. The relative coupling can then be chosen so to absorb part of the divergences  
and tune the renormalization of the abelian fields with the one for the non-abelians. In so doing, 
a renormalizable, $N=1/2$ and gauge invariant cubic superpotential can be added. 
On the other side, it changes the gauge--matter coupling in vertices where abelian (anti)chirals 
are present. 
As a crucial consequence, the evaluation of one--loop diagrams reveals that only $N=1/2$ susy and 
supergauge invariant divergent structures get produced. 
Therefore, a one--loop renormalizable action is obtained by 
adding all possible $N=1/2$ supergauge invariant couplings allowed by dimensional analysis. 
Its explicit expression is given in eq. (\ref{main}). 

In this paper we construct the renormalizable action by 
performing a dimensional and diagrammatic analysis of divergences, without entering the details of the 
calculations. 
The paper is organized as follows: In Section 2 we review the NAC superspace and the generalization
of the background field method already discussed in \cite{GPR2}. In Sections 3 and 4 we discuss the 
quantization of the NAC SYM model obtained by promoting ordinary products to be $\ast$--products in 
the action for $SU({\cal N}) \times U(1)$ gauge superfields coupled to chiral matter in the adjoint 
representation, in the presence of a cubic superpotential. We formulate 
the superpotential problem in the language of superspace and propose our solution which requires 
introducing a different coupling constant in front of the abelian quadratic action. In Section 5 we perform
a general selection of all possible divergent structures which might appear at loop level and propose 
the most general one--loop renormalizable gauge--invariant action. Finally, in Section 6 we prove 
that all divergences can be multiplicatively renormalized while preserving gauge--invariance.
Finally, Section 7 contains some conclusions and perspectives. An Appendix follows where we derive
the Feynman rules necessary for perturbative calculations.

\section{The general setting}

We consider the $N=(\frac12 , 0)$ NAC superspace spanned by nonanticommutative coordinates 
$(x^{\a \ad}, \th^\a, \thb^\ad)$ satisfying
\beq
\{ \th^\a, \th^\b \} = 2 {\cal F}^{\a\b} \qquad \{ \thb^\ad, \thb^\bd \} = 0 \qquad [x^{\a\ad}, x^{\b \bd}] = 
[ x^{\a\ad}, \th^\b] = [ x^{\a\ad}, \thb^\bd] =0
\eeq
where ${\cal F}^{\alpha \beta}$ is a $2 \times 2$ symmetric, constant matrix. This algebra is
consistent only in euclidean signature where the chiral and antichiral sectors
are totally independent and not related by complex conjugation.

The class of smooth superfunctions on the NAC superspace is endowed with the 
NAC but associative product 
\beq
\phi \ast \psi ~\equiv~ \phi e^{- \overleftarrow{\pa}_\a {\cal F}^{\a \b}
\overrightarrow{\pa}_\b} \psi
~=~ \phi \psi - \phi \overleftarrow{\pa}_\a {\cal F}^{\a \b}
\overrightarrow{\pa}_\b \psi - \frac12 {\cal F}^2 \pa^2\phi \, {\pa}^2 \psi
\label{star} 
\eeq
where ${\cal F}^2 \equiv  {\cal F}^{\a \b} {\cal F}_{\a \b}$. 
(Anti)chiral superfields can be consistently defined by the constraints 
$\overline{D}_{\ad} \ast \phi = D_\a \ast \overline{\phi} =0$ \footnote{We use chiral 
representation \cite{superspace} for supercharges and covariant derivatives. In particular, 
we define $D_{\a} = \pa_{\a} + i {\overline{\theta}}^{\ad} \pa_{\a \ad}$ and 
$\overline{D}_{\ad} =  \overline{\pa}_{\ad}$.}.

\vskip 15pt

Supersymmetric Yang-Mills theories in NAC superspace have been extensively discussed in Ref. \cite{GPR2}.
As in the ordinary case, they are defined in terms of a scalar prepotential $V\equiv V_A T^A$ 
in the adjoint representation of the gauge group. Being the theory in euclidean signature, 
$V$ has to be {\em pure imaginary}, $V^\dag = -V$.

The supergauge transformations for $V$ are
\beq
e_\ast^V \rightarrow  e_\ast^{V'} = e_\ast^{i \overline{\L}} \ast
e_\ast^V \ast e_\ast^{-i\L}
\label{gauge}
\eeq
where  $\L, \overline{\L}$ are chiral and antichiral superfields, respectively.

Supergauge covariant derivatives in superspace can be defined in the so--called
{\em gauge chiral} or {\em gauge antichiral} representation \cite{superspace}. As discussed 
in \cite{GPR2} in the NAC case the two descriptions are no longer equivalent, especially
when the construction of supergauge invariant actions is under concern. It turns out
that the {\em gauge antichiral} representation is definitely preferable. We then define
supergauge covariant derivatives as 
\beq
{\nabla}_A \equiv (\nabla_\a , \overline{\nabla}_{\ad},
\overline{\nabla}_{\a \ad})
~=~ (D_\a ~,~ e_\ast^{V} \ast \overline{D}_{\ad} \, e_\ast^{-V} ~,~
-i \{ \nabla_\a, \overline{\nabla}_{\ad} \}_{\ast} )
\label{derivatives}
\eeq
They can be expressed in terms of ordinary superspace derivatives and a set of connections, 
${\nabla}_A \equiv D_A - i \G_A$, where
\beq
{\G}_{\a}=0     \qquad , \qquad  \overline{\G}_{\ad} =
 ie_\ast^{V} \ast \overline{D}_{\ad} \, e_\ast^{-V}
\qquad , \qquad \overline{\G}_{\a \ad} = -i D_\a  \overline{\G}_{\ad}
\eeq
The field strengths are then defined as $\ast$--commutators of supergauge
covariant derivatives
\beq
\overline{W}_{\ad} = -\frac12 [ \nabla^{\a},
\overline{\nabla}_{\a \ad} ]_\ast \qquad , \qquad 
\Tilde{W}_\a =
-\frac12 [ \overline{\nabla}^{\ad}, \overline{\nabla}_{\a \ad} ]_\ast
\eeq
and satisfy the Bianchi's identities $\nabla^\a \ast \Tilde{W}_\a + 
\overline{\nabla}^{\ad} \ast \overline{W}_{\ad}=0$.
In terms of gauge connections they are given by
\beq
\overline{W}_\ad = \frac{i}{2} D^\a \overline{\G}_{\a \ad} = D^2 \overline{\G}_\ad
\qquad , \qquad \Tilde{W}_\a = \frac{i}{2} \pa_\a^{\, \ad} \overline{\G}_\ad 
+ \frac{i}{2} [\overline{\nabla}^\ad , \overline{\G}_{\a \ad}]_\ast
\label{fieldstrengths}
\eeq
Covariantly (anti)chiral superfields can be defined according to 
$\overline{\nabla}_\ad \ast \Phi = 0$ and $\nabla_\a \ast \overline{\Phi} = 0$, 
respectively.

At classical level, a NAC SYM theory with interacting chiral matter in 
the adjoint representation of $SU(\Nc) \otimes U(1)$ can be described by the following
action \cite{GPR2}
\bea
\label{classaction}
&& S =
\frac{1}{2g^2} \int d^4x d^2 \thb ~{\rm Tr} (\overline{W}^{\ad}
\ast \overline{W}_{\ad})
\\
&&~~~+ \frac{1}{2g_0^2}\int d^4x~d^4 \theta~\Big[ ~{\rm Tr}( \overline{\G}^{\ad})
\ast {\rm Tr}(\overline{W}_{\ad})
+ 4 i {\cal F}^{\rho \g} \overline{\theta}^2~{\rm Tr} \left( \pa_{\rho \dot{\rho}}
\overline{\Gamma}^{\ad} \right) \ast {\rm Tr}\left(
\overline{W}_{\ad} \ast \overline{\Gamma}_{\g}^{~\dot{\rho}} \right) \Big]
\nonumber \\
&&~~~+ \int d^4x~d^4 \theta ~{\rm Tr} (\overline{\Phi} \ast \Phi)
\non \\
&&~~~+ h \int d^4x~d^2 \theta ~{\rm Tr} (\Phi \ast \Phi \ast \Phi)  
+ \bar{h} \int d^4x~d^2 \thb ~{\rm Tr} (\overline{\Phi} \ast \overline{\Phi} \ast 
\overline{\Phi}) \non
\eea
where $\Phi \equiv e^V_\ast  \ast \phi \ast e^{-V}_\ast$, $\Phib = \overline{\phi}$ 
are covariantly (anti)chiral superfields expressed in terms of ordinary (anti)chirals. 
Therefore, the quadratic matter action contains nontrivial couplings between gauge and chiral 
superfields.

The action is invariant under the infinitesimal supergauge transformations
\bea
\d \Phi = i [ \overline{\L} , \Phi]_\ast \qquad &,& \qquad
\d \overline{\Phi} = i [ \overline{\L}, \overline{\Phi}]_\ast
\nonumber \\
\d \overline{\G}_{\a\ad} =  [\overline{\nabla}_{\a\ad},  \overline{\L} ]_\ast \qquad &,& \qquad
\d \overline{W}_{\ad} = i [ \overline{\L} , \overline{W}_{\ad} ]_\ast
\label{Wtransf}
\eea
As discussed in \cite{GPR2} the term proportional to $\thb^2$ in (\ref{classaction}) is necessary 
in order to 
restore gauge invariance of  $\int {\rm Tr}( \overline{\G}^{\ad}){\rm Tr}(\overline{W}_{\ad})$.

The transformation law for $\overline{W}_\ad$ can be rewritten as
\beq
\d \overline{W}_\ad^A = \frac{i}{2} d_{ABC} [ \bar{\L}^B , \Wbar_\ad^C ]_\ast - \frac12  f_{ABC}
\{ \bar{\L}^B , \Wbar_\ad^C \}_\ast
\label{Wtransf2}
\eeq
where $A,B,C$ are $SU({\cal N}) \otimes U(1)$ indices. The first term  
is non-vanishing only in the NAC case and mixes nontrivially $U(1)$ and $SU({\cal N})$ fields. 
In particular, the abelian field strength $\Wb_\ad^0$ is no longer a singlet but transforms 
under $SU({\cal N})$ into a linear combination of both $U(1)$ and $SU({\cal N})$ fields. 

\vskip 15pt

In superspace, a convenient procedure for performing perturbative calculations for 
super Yang--Mills theories is the background field method \cite{GSZ, superspace}.
It consists of  a nonlinear
quantum--background  splitting on the gauge superfields which leads to separate
background  and quantum gauge invariances. Gauge fixing is then chosen which breaks the 
quantum  invariance while keeping manifest invariance with respect to the background gauge 
transformations. Therefore, at any given order in the loop expansion the contributions to the
effective action are expressed directly in terms of covariant derivatives and
field strengths (without  explicit dependence on the prepotential $V$). 

The generalization of the background field method to NAC SYM theories with chiral matter in a 
{\em real} representation of the gauge group has been performed in \cite{GPR2}. 
Here we summarize the main ingredients referring the reader to that paper for details. 

We perform the splitting  of the Euclidean prepotential $e^V _\ast
\rightarrow e^V_\ast \ast e^{U}_\ast$ where $U$ is the background prepotential and $V$ its 
quantum counterpart. 
Consequently, the covariant derivatives (\ref{derivatives}) become
\beq
\nabla_\a = \boldnabla_\a =D_\a \qquad , \qquad \overline{\nabla}_{\ad} = 
e_{\ast}^V \ast \overline{\boldnabla}_{\ad}\ast e_{\ast}^{-V} = e_{\ast}^V \ast
(e_{\ast}^U\ast\Db_{\ad} \ e_{\ast}^{-U}) \ast e_{\ast}^{-V}
\label{bqsplitting}
\eeq
Covariantly (anti)chiral superfields in the adjoint representation are expressed in terms
of background covariantly (anti)chiral objects as 
\beq
\overline{\Phi} =
 \bold{\overline{\Phi}} \qquad \qquad , \qquad \qquad
   \Phi = e^V_{\ast} \ast \bold{\Phi}\ast e^{-V}_\ast
   =e_\ast^V \ast (e_\ast^U \ast \phi \ast e_\ast^{-U}) \ast e^{-V}_\ast
   \label{covchiral}
\eeq
and then splitted as $ \bold{\Phi} \to  \bold{\Phi} +  \bold{\Phi}_q$ and 
$ \bold{\Phib} \to  \bold{\Phib} +  \bold{\Phib}_q$, where  $\bold{\Phi}, \bold{\Phib}$ are
background fields and $\bold{\Phi}_q, \bold{\Phib}_q$ their quantum fluctuations. 

We perform quantum-background splitting in the action (\ref{classaction}) and
extract the Feynman rules necessary for one--loop calculations.

\vskip 10pt
\noindent
\underline{Gauge sector}

As in the ordinary case, the invariance under quantum gauge transformations 
\cite{GSZ, superspace,GPR2} is broken by choosing gauge--fixing functions as 
$f = \overline{\boldnabla}^2 \ast V$, $\overline{f} = {\boldnabla}^2 \ast V$, 
while preserving manifest invariance of the effective action and correlation
functions under background gauge transformations \cite{GSZ, superspace,GPR2}.
In Ref. \cite{GPR2} the gauge--fixing procedure for $SU(\Nc) \otimes U(1)$ NAC
gauge theories has been discussed in detail. 
As a result, extracting the quadratic part in the quantum 
$V$ fields from $\int d^4x d^2 \thb [\frac{1}{2g^2} \Tr (\Wbar^\ad \Wbar_\ad) + 
\frac{1}{2g_0^2} \Tr (\Wbar^\ad )\Tr(\Wbar_\ad) ]$ and adding the gauge--fixing action
\beq
S_{GF} = - \frac{1}{g^2 \a} \int d^4x d^4 \th \, \Tr \left[ (\overline{\boldnabla}^2 \ast V) 
(\boldnabla^2 \ast V) \right] 
\label{gf}
\eeq
in Feynman gauge we find
\bea
S \rightarrow &-& \frac{1}{2g^2} \int d^4 x d^4 \th  ~ \left[ V^a \ast \hat{\Box} \ast V^a 
\right.
\\
&~&~~~~~~ \left. + V^0 \ast \Big( \boldnabla^2 \ast \overline{\boldnabla}^2 
+ \overline{\boldnabla}^2 \ast \boldnabla^2 - \frac{g_0^2+g^2}{g_0^2} \overline{\boldnabla}^\ad
\ast \boldnabla^2 \ast \overline{\boldnabla}_\ad \Big) \ast V^0 \right]  
\non
\eea
where the label $a$ runs over $SU(\Nc)$ indices and we have defined
\beq
\hat{\Box} = \Box_{cov} - i\widetilde{\bold{W}}^\a \ast \boldnabla_\a - i \overline{\bold W}^\ad 
\ast \overline{\boldnabla}_\ad
\qquad , \qquad \Box_{cov} = \frac12  \overline{\boldnabla}^{\a\ad} \ast 
\overline{\boldnabla}_{\a\ad}
\label{boxhat}
\eeq
Introducing also the covariant operator
\beq
\label{tildebox}
\Tilde{\Box} = \boldnabla^2 \ast \overline{\boldnabla}^2 + \overline{\boldnabla}^2 \ast 
\boldnabla^2 - \overline{\boldnabla}^\ad \ast \boldnabla^2 
\ast \overline{\boldnabla}_\ad = \Box_{cov} - i \Tilde{\bold W}^\a \ast \boldnabla_\a 
+ \frac{i}{2} 
( \overline{\boldnabla}^\ad \ast \Wbar_\ad) 
\eeq
perturbative contributions can be written in terms of background covariant propagators 
\bea
\langle V^a V^b \rangle &=& g^2\left( \frac{1}{\hat{\Box}} \right)^{ab}
\non \\
\langle V^0 V^0 \rangle &=&  g^2 \left\{ \frac{1}{\Tilde{\Box}} \left[ 1 + \left( \frac{g^2}{g^2 + g_0^2} 
\right) \overline{\boldnabla}^\ad \ast \boldnabla^2 \ast \overline{\boldnabla}_\ad \ast
\frac{1}{\Tilde{\Box}} \right]\right\}^{00}
\eea
Their expansion in powers of the background fields provides the ordinary $\frac{1}{\Box}$ 
propagator for abelian and non--abelian superfields plus pure-gauge interaction vertices (see
Appendix A). 

Further vertices come from the 
background field expansion of the $\thb^2$ term in the second line of action (\ref{classaction}). 
Their explicit expressions can be found in Appendix E of Ref. \cite{GPR2}.

The ghost action associated to the gauge--fixing (\ref{gf}) is given in terms of background 
covariantly (anti)chiral FP and NK ghost superfields as
\beq
S_{gh} =  \int d^4x d^4 \theta ~ \Big[ \overline{c}' c - c'\overline{c} + .....+ \overline{b} b \Big]
\label{ghosts}
\eeq

\vskip 10pt
\noindent
\underline{Chiral sector}

We now discuss the quantization of the matter action in (\ref{classaction}) when $\Phi, \Phib$ are full 
convariantly (anti)chiral superfields. With obvious modifications the results hold also for the background covariantly 
chiral ghosts in (\ref{ghosts}). 

We first express the full covariantly (anti)chiral superfields in terms of background covariantly (anti)chiral 
superfields according to (\ref{covchiral}). Expanding in powers of $V$ we have 
(we use the notation $\Phi_\ast^3 \equiv \Phi \ast \Phi \ast \Phi$)
\bea
S_0 + S_{int} = \intsup {\bf \Phib} \ast {\bf\Phi} 
&+& \intsup \left( {\bf \Phib} [V, {\bf\Phi}]_\ast + \frac{1}{2} {\bf\Phib} [V,
    [V,{\bf\Phi}]_\ast]_\ast + \ldots \right) 
\non \\
&+& h \intch \Tr( \bold{\Phi}^3_\ast) + \overline h \intach \Tr( \bold{\Phib}^3_\ast )  
\label{chiralaction}
\eea
where the Trace over group indices has been omitted since the quantization procedure works 
independently of the color structure.   
After the shift $\bold{\Phi} \to \bold{\Phi} + \bold{\Phi}_q$,
$\bold{\Phib} \to \bold{\Phib} + \bold{\Phib}_q$ only terms with two quantum 
superfields need be considered for one--loop calculations.  

Quantization is accomplished by adding source terms 
\bea
S_j &=& \intch j \ast \bold{\Phi}_q + \intach \bold{\Phib}_q \ast \overline j 
\non \\
&=& \intsup
\left( j \ast \frac{1}{\Box_+} \ast \boldnabla^2 \bold{\Phi}_q + \bold{\Phib}_q \ast \frac{1}{\Box_-}
\ast \overline{\boldnabla}^2 \ast \overline j \right)
\eea
where, for any (anti)chiral superfield, we have defined 
\bea
&& \Delb^2 \ast \Del^2 \ast \Phi= \Box_+  \ast \Phi
\qquad \quad \quad
\Box_+ = \Box_{cov} - i \widetilde{W}^\a \ast \nabla_\a
    -\frac{i}{2}(\nabla^\a \ast \widetilde{W}_\a)
\non \\
&& \Del^2  \ast \Delb^2 \ast \Phib = \Box_- \ast \Phib
\qquad \quad \quad 
\Box_- = \Box_{cov} - i
    \Wbar^\ad \ast \Delb_\ad -\frac{i}{2}(\Delb^\ad \ast \Wbar_\ad)
    \non \\
    &&
\label{boxes}
\eea
and performing the gaussian integral in 
\beq
  Z = \int {\cal D}\bold{\Phi}_q {\cal D}\bold{\Phib}_q
  e^{S_{int}(\frac{\delta}{\delta j}, \frac{\delta}{\delta \overline
      j})} e^{\intsup (\bold{\Phib}_q \ast \bold{\Phi}_q + j \ast \frac{1}{\Box_+} \ast \boldnabla^2 
      \bold{\Phi}_q
    + \bold{\Phib}_q \ast \frac{1}{\Box_-} \ast \overline{\boldnabla}^2 \ast \overline j)}
\label{effective0}
\eeq
The Feynman rules can then be read from 
\beq
  Z = \Delta ~e^{S_{int}(\frac{\delta}{\delta j},
    \frac{\delta}{\delta \overline j})} 
    e^{-  \intsup   j \ast 
    \frac{1}{\Box_-} \ast \overline j }
\label{effective}
\eeq
where $\Delta \equiv \int {\cal D}\bold{\Phi}_q {\cal D}\bold{\Phib}_q e^{S_{0}}$. In particular, we obtain
the covariant scalar propagator
\beq
\langle \Phi^A \Phib^B \rangle = - \left( \frac{1}{\Box_-} \right)^{AB}
\eeq

At one--loop, from the matter sector we have two different contributions to the effective action.
A first contribution to the gauge effective action comes from the perturbative evaluation of $\Delta$.  
This can be worked out by using the doubling trick procedure introduced in \cite{superspace} for ordinary 
SYM theories and generalized to NAC theories in \cite{GPR2}. The corresponding Feynman rules are collected
in Ref. \cite{GPR2}.
A second contribution comes from the perturbative expansion of $e^{S_{int}}$ from which we can read 
gauge--chiral vertices. Further interaction vertices arise from the expansion of  $1/\Box_-$ in powers of the background fields (see Appendix A).

\section{One--loop divergences: The gauge sector}

In Ref. \cite{GPR2} we computed divergent contributions to the pure gauge sector of 
the NAC $SU({\cal N}) \otimes U(1)$ SYM theory. 
It turned out that the classical action (\ref{classaction}) is not renormalizable since 
further divergent configurations arise at one--loop which are $N=1/2$ supersymmetric and supergauge
invariant. However, we proved that it is possible to deform the classical action in such a way as to
produce a one--loop renormalizable theory. The manner in which we proceeded is to
start {\em ab initio} with a deformed action containing all possible terms allowed by
gauge invariance, R--symmetry, and dimensional analysis. We computed all one--loop divergences 
produced by the new action and determined a one-loop renormalizable action depending on a number of
arbitrary coupling constants. Computing the $\b$-functions we found that
they allow for specific restrictions on these constants.
In particular, two different choices for minimal deformed actions are allowed which are
one-loop renormalizable 
\bea
&& S_{gauge}^{(1)} =
~\frac{1}{2~g^2}\int d^4x~d^4\theta~ {\rm Tr}\left(
\overline{\Gamma}^{\ad} \ast \overline{W}_{\ad}\right) \nonumber \\
&& \qquad
+ \frac{1}{2~g_0^2~{\cal N}} \int d^4x~d^4 \theta~ \Bigg[ ~~ {\rm
Tr}\left( \overline{\Gamma}^{\ad} \right)\ast {\rm Tr} \left(
\overline{W}_{\ad}\right) \nonumber \\
&& \qquad \qquad  \qquad \qquad \qquad\qquad + 4 i {\cal
F}^{\rho \g} \overline{\theta}^2~{\rm Tr} \left(
\pa_{\rho \dot{\rho}} \overline{\Gamma}^{\ad} \right) \ast {\rm Tr}\left(
\overline{W}_{\ad} \ast \overline{\Gamma}_{\g}^{~\dot{\rho}} \right)
 \nonumber \\
&& \qquad \qquad \qquad \qquad \qquad \qquad - {\cal F}^2
\overline{\theta}^2~{\rm Tr}
\left(\overline{\Gamma}^{\ad} \ast \overline{W}_{\ad}\right){\rm
Tr}\left(\overline{W}^{\bd} \ast \overline{W}_{\bd}\right) ~~~\Bigg] \nonumber \\
&& \qquad + \frac{1}{l^2} ~{\cal F}^2 \int d^4 x ~d^4
\theta~\overline{\theta}^2 ~{\rm Tr} \left( \overline{\Gamma}^{\ad} \ast
\overline{W}_{\ad} \ast \overline{W}^{\bd} \ast \overline{W}_{\bd}\right)
\label{min1}
\end{eqnarray}
or
\bea
&& S_{gauge}^{(2)} =
~\frac{1}{2~g^2} \int d^4 x ~ d^4 \theta~ \Bigg[ ~~{\rm Tr}\left(
\overline{\Gamma}^{\ad} \ast \overline{W}_{\ad}\right) \nonumber \\
&& \qquad  \qquad \qquad \qquad  \qquad \qquad+ {\cal F}^2
~ \overline{\theta}^2~{\rm Tr}
\left(\overline{\Gamma}^{\ad} \ast \overline{W}_{\ad}\right) \ast {\rm
Tr}\left(\overline{W}^{\bd} \ast \overline{W}_{\bd}\right) \Bigg] \nonumber \\
&& \qquad
+ \frac{1}{2~g_0^2~{\cal N}} \int d^4x~d^4 \theta~ \Bigg[ ~~{\rm
Tr}\left( \overline{\Gamma}^{\ad} \right) \ast {\rm Tr}\left(
\overline{W}_{\ad}\right) \nonumber \\
&& \qquad  \qquad \qquad \qquad  \qquad \qquad+ 4 i {\cal
F}^{\rho \g} ~ \overline{\theta}^2~{\rm Tr} \left(
\pa_{\rho \dot{\rho}} \overline{\Gamma}^{\ad} \right) \ast {\rm Tr}\left(
\overline{W}_{\ad} \ast \overline{\Gamma}_{\g}^{~\dot{\rho}} \right) \Bigg]
 \nonumber \\
&& \qquad + \frac{1}{l^2} ~{\cal F}^2 \int d^4 x ~d^4
\theta~\overline{\theta}^2 ~{\rm Tr} \left( \overline{\Gamma}^{\ad} \ast
\overline{W}_{\ad} \ast \overline{W}^{\bd} \ast \overline{W}_{\bd}\right)
\label{min2}
\end{eqnarray}
In both cases the theory contains three independent coupling constants. While $g, g_0$ are 
the $SU({\cal N}) \otimes U(1)$ couplings already present in the ordinary theory, the appearance
of the third coupling $l$ is strictly related to the NAC deformation we have performed. We note that 
$g, g_0$ must be different reflecting the fact that, as in the ordinary case, $SU({\cal N})$ and
$U(1)$ fields renormalize differently.

\section{One--loop divergences: The matter sector}

We now study the structure of one--loop divergent contributions to the matter sector with 
particular attention to the superpotential problem. 

We begin with the classical action 
\beq
S_{matter} = \int d^4x~d^4 \theta ~{\rm Tr} (\overline{\Phi} \ast  \Phi) 
+ h \int d^4x~d^2 \theta ~{\rm Tr} (\Phi_\ast^3)  
+ \bar{h} \int d^4x~d^2 \thb ~{\rm Tr} (\overline{\Phi}_\ast^3)
\label{classaction2}
\eeq
for covariantly (anti)chiral superfields. Applying background field method 
we evaluate one--loop diagrams with external matter.

\subsection{The quadratic action}

\begin{figure}
  \begin{center}
    \includegraphics[width=0.25\textwidth]{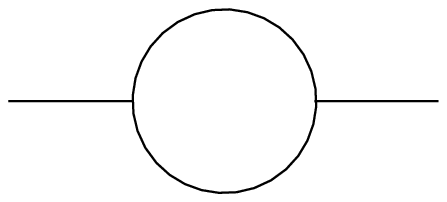}
    \includegraphics[width=0.25\textwidth]{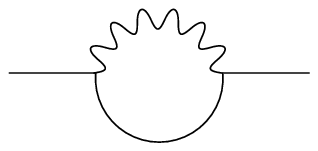}
  \end{center}
  \caption{One--loop two--point functions with chiral external
    fields. }
  \label{fig:chiral0}
\end{figure}

Divergent diagrams contributing to the two--point function are given in Fig. \ref{fig:chiral0} where 
the internal lines correspond to ordinary $1/\Box$ propagators 
(straight lines correspond to chiral propagators, whereas waved lines correspond to vectors). 
It turns out that divergent contributions to the quadratic term come only from vertices not 
including the deformation parameter. Therefore, they coincide with the ones of the underformed theory 
and are given by  
\beq
\mathcal{S} \intsup \Big[ \left( 9h\overline{h} - 2g^2  \right) \mathN ~\Tr \left( \bold{\Phib}
\ast \bold{\Phi} \right) 
+ \left( 9h\overline{h} + 2g^2 \right)  ~\Tr \bold{\Phib}  \ast \Tr \bold{\Phi} \Big]
\label{2point}
\eeq
where $\mathcal{S}$ is the self--energy divergent integral which in dimensional regularization 
is ($d= 4-2\e$)  
\bea
  {\cal S} \equiv
\int d^d q ~\frac{1}{q^2(q-p)^2}
= \frac{1}{(4\pi)^2}~\frac{1}{\epsilon} + {\cal O}(1)
\label{selfenergy}
\eea

We note that a new trace structure appears reflecting the fact that $SU({\cal N})$ and
$U(1)$ superfields acquire different contributions. In fact, considering only the kinetic term, 
the previous result reads (using eq. (\ref{covchiral}))
 \beq
\mathcal{S} \intsup \left[ \left( 9h\overline{h} - 2g^2  \right) \mathN 
\phib^a \phi^a  + 18 h\overline{h} \mathN \phib^0  \phi^0 \right]
\eeq
In particular, corrections to the abelian kinetic term coming from the gauge--chiral loop cancel 
in agreement with the calculation done in components \cite{JJW3}. 

In the ordinary case, the appearance of the double--trace term is harmless since it is supergauge
invariant. In the NAC case this is no longer true since its variation is
\beq
\delta\, \Tr \Phibbold \ast \Tr \Phibold = 2i \thb^2 \Fab \left[ \Tr \left( \partial_\alpha^{~\ad} 
\Lb \ast \partial_{\beta\ad} \Phibbold \right) \ast  \Tr \Phibold + \Tr \Phibbold \ast \Tr \left(
\partial_\alpha^{~\ad} \,\Lb \ast \partial_{\beta\ad} \, \Phibold \right) \right]
\eeq
and does not vanish when integrated on superspace coordinates.

On general grounds, it is easy to see that there are two possible gauge completions for 
$\int\Tr \Phibbold \ast \Tr \Phibold$. In fact,
the following expressions (both for background covariantly and full covariantly (anti)chiral
superfields) 
\beq
\Tr \Phib \ast \Tr \Phi + 2i \thb^2 \Fab~ \Tr
  ( \Gbar_{\a}^{~ \ad} \ast \Phib ) \ast \Tr (
  \partial_{\beta \ad} \Phi )
+ 2i \thb^2 \Fab~ \Tr ( \Gbar_{\alpha}^{~ \ad} \ast \Phi) 
\ast \Tr ( \partial_{\beta \ad} \Phib )
\label{gauge1}
\eeq
and
\bea
\label{gauge2}
\Tr \Phib \ast \Tr \Phi &-& 2i\Fab \thb^2  \Tr
\left[ \Gbar_{\a}^{~ \ad} \ast \left( \partial_{\beta \ad} \Phib 
-\frac{i}{2} [\Gbar_{\beta\ad}, \Phib]_\ast \right) \right] \ast \Tr \left( \Phi \right)
\non  \\ 
&-& 2i\Fab \thb^2 \Tr \left[
    \Gbar_{\alpha}^{~ \ad} \ast \left( \partial_{\beta \ad} \Phi
    -\frac{i}{2} [\Gbar_{\beta\ad}, \Phi]_\ast \right) \right] \ast \Tr
  \left( \Phib \right) 
\eea
are both gauge invariant when integrated.
While the first expression involves only gauge--chiral cubic terms in
addition to the quadratic term, the second one involves also quartic couplings. 
Therefore, we have to investigate whether at one--loop the theory develops further divergent 
terms cubic and/or quartic in the background fields 
which provide the gauge completion of $\int \Tr \Phibold \ast \Tr \Phibbold$.

Divergences proportional to gauge--chiral cubic terms are still
obtained from diagrams in Fig. \ref{fig:chiral0} where the internal lines correspond to covariant
$1/\Box_-$ and $1/\hat{\Box}$ propagators expanded up to quadratic order 
in the background gauge superfields (see eqs. (\ref{eqn:box_{vector}}, \ref{eqn:box_{cov}})). 
Summing the contributions coming from both diagrams in Fig. \ref{fig:chiral0} we obtain
\bea
\label{3pt}
&& (9 h \overline h + 2g^2) ~{\cal S}  \intsup 2i ~\Fab \thb^2 \left[
\Tr (\Gbar_\a^{~\ad} \ast \Phibbold) \ast \Tr ( \partial_{\beta\ad} \Phibold )
+ \Tr ( \Gbar_\alpha^{~\ad} \ast \Phibold) \ast \Tr ( \partial_{\beta\ad}
\Phibbold ) \right]
\non \\
&~& ~~~~~~~~-2i ( 9 h \overline h - 2g^2) ~{\cal S} 
\intsup ~\Fab \thb^2 \Tr ( \partial_{\beta\ad} \Gbar_\alpha^{~\ad} ) \ast \Tr ( \Phibold \ast\Phibbold ) 
\eea
The first line is exactly the gauge completion of (\ref{2point}) according to (\ref{gauge1}). 
In addition, a second divergent term appears in the second line. 
Since it is gauge invariant it is allowed by super Ward identities. 

We should not expect divergent four--point functions proportional to $\Gbar_{\a\ad}$ 
connections since there is no need to saturate gauge--variation of two--point divergences. 
In fact, from a direct inspection one can realize that only structures of the form
\bea
\label{4pt}
{\cal F}^2~ \intsup
\thb^2~\Phibold \ast \Phibbold \ast \Wb^{\ad} \ast \Wb_{\ad}
\eea
can be divergent. For any kind of trace structure all
these terms are gauge--invariant and do not interfere with the previous structures.

\vskip 20pt

To summarize, the evaluation of one--loop divergences reveals that
the action (\ref{classaction2}) we started with is not renormalizable because of the 
appearance of new one--loop structures not originally present. 

At this stage
it is easy to generalize the classical action to a renormalizable one in a gauge invariant way: 
It is sufficient to start with a classical quadratic action of the form 
\bea
\label{2pointdiv}
&&\intsup ~\Big\{ \Tr \left( \Phib \ast \Phi \right)
\\
&&+ \Big[ \Tr \Phib  \ast \Tr 
\Phi + 2i \thb^2 \Fab \Tr (\Gbar_\a^{~\ad} \ast \Phib) \ast \Tr ( \partial_{\beta\ad} \Phi )
+ 2i \thb^2 \Fab \Tr ( \Gbar_\alpha^{~\ad} \ast \Phi) \ast \Tr ( \partial_{\beta\ad} \Phib ) \Big]
\Big\}
\non
\eea
supplemented by the gauge invariant terms appearing in 
the second line of (\ref{3pt}) and in (\ref{4pt}). 

We stress once again that the divergent contributions (\ref{2point}) 
to the quadratic action would be present
also in the ordinary, not deformed theory. Therefore, also in that case we would be forced to 
generalize the classical quadratic action to contain a double--trace part, in order to make the
theory renormalizable. The crucial difference is that the double--trace term would be gauge invariant 
and no gauge completion would be required. 

As already mentioned, in the NAC case the double trace quadratic action has in principle two possible gauge completions. 
From direct inspection, the theory seems to prefer the gauge invariant structure (\ref{gauge1}) 
rather than (\ref{gauge2}).

\subsection{The superpotential problem}

As we now describe, when a chiral superpotential is turned on the generalization (\ref{2pointdiv})
for the quadratic matter action is not sufficient to make the theory renormalizable. 
 
Since the nonrenormalization theorem for chiral integrals works also in the NAC case \cite{BFR2,GPR1},
the cubic superpotential in (\ref{classaction2}) does not get corrected by new diagrams proportional 
to $\Phi^3$ and/or $\Phib^3$. As in the ordinary case, the renormalization of the 
chiral coupling constant is induced by the wave--function renormalization under the requirement that
$Z_h Z_{\Phi}^{-3/2} = 1$ (a similar relation holds for the antichiral coupling). 
On the other hand, $SU(\mathN)$ and $U(1)$ 
chiral superfields renormalize differently, so should do the corresponding chiral couplings. Therefore,
a cubic superpotential as the one in (\ref{classaction2}) which assigns the same coupling to 
the $SU(\mathN)$, $U(1)$ and mixed interaction vertices is inconsistent with the request of 
renormalizability. 
We note that this problem 
is not peculiar of the NAC deformation being present already in the ordinary case.

The way out is once again the generalization of the classical action to include different couplings for
different cubic vertices. Exploiting the fact that in Euclidean space $Z_{\Phib}$ is not necessarily equal to 
$Z_{\Phi}$, we can trigger the renormalization in such a way 
that for instance all the renormalization asymmetry  between non--abelian and abelian fields is 
confined to the antichiral sector.  As a consequence, we can consistently choose the ordinary $h \int d^4x~d^2 \th
 \Tr (\Phi_\ast^3)$ superpotential in the chiral sector, but generalize the one for the antichiral sector
 to
 \beq
\int d^4x~d^2 \thb ~\Big[ \bar{h}_1 \Tr (\Phib_\ast^3 )  + 
\bar{h}_2 \Tr \Phib \ast \Tr (\Phib_\ast^2) + \bar{h}_3 (\Tr \Phib)_\ast^3 \Big]
\label{generalized} 
\eeq
However, while in the ordinary case the different structures are separately gauge invariant, 
in the NAC case the addition of the $\bar{h}_2, \bar{h}_3$ terms breaks gauge invariance. In fact, 
due to the lack of $\th$--integration, the traces are no longer cyclic and  
$\d \int (\Tr \Phib  \ast \Tr (\Phib_\ast^2) )$ and $\d \int (\Tr \Phib)_\ast^3$ are non--vanishing.

The gauge completion of these terms reads
\bea
    &\bar{h}_2 \intach~\Bigg\{
    \Tr\Big( \Phib -2i \thb^2 \Fab~ \Gbar_{\alpha}^{~\ad} *
    \left\{ \partial_{\beta \ad} \Phib - \frac{i}{2} \left[
      \Gbar_{\beta \ad}, \Phib \right]_* \right\} \Big) * \Tr
    \left( \Phib^2_{\ast} \right) 
    \non \\ 
    &~~~~~~+ \Tr \Phib *
    \Tr \Big( \Phib^2_{\ast} -2i \thb^2 \Fab~ \Gbar_{\alpha}^{~\ad}
    * \left\{ \partial_{\beta \ad} \Phib^2_{\ast} - \frac{i}{2} \left[
      \Gbar_{\beta \ad}, \Phib^2_{\ast} \right]_* \right\} \Big
      )
    \Bigg\}  
\eea
and
\beq
\bar{h}_3 \intach\! \Tr\! \left(\! \Phib -6i \thb^2 \Fab
    \Gbar_{\alpha}^{~\ad}\! *\! \left\{\! \partial_{\beta \ad}
    \Phib - \frac{i}{2} \left[ \Gbar_{\beta \ad}, \Phib
      \right]_*\!  \right\}\! \right)\! *\! \Tr\! \left( \Phib
    \right)\! *\! \Tr\! \left( \Phib \right) 
\eeq
respectively.

The terms proportional to $\Gbar_{\a\ad}$ in the previous expressions break supersymmetry
completely since they are given by non--antichiral expressions integrated over an antichiral measure. 
Therefore, one--loop renormalizability, gauge invariance and $N=1/2$ supersymmetry seem to be 
incompatible.
This is the translation in superspace language of the negative result already found in 
components \cite{JJW3}.

\subsection{The solution to the superpotential problem}

Fortunately, generalizing the superpotential to contain more than one coupling constant does not seem 
to be the only possibility for constructing a renormalizable action.  
In fact, an alternative procedure exists for treating the diverse renormalization of the abelian 
fields in a consistent way. The idea is to start with a classical quadratic action of the form (\ref{2pointdiv}) 
but with a new coupling in front of the double--trace term 
\bea
\label{quadratickappa}
&& \intsup  \Big\{ \Tr \left( \Phib \ast \Phi \right)
+ \frac{\kappa - 1}{\Nc} \Big[ \Tr \Phib  \ast \Tr \Phi 
\\
&~&~~~\quad ~\quad ~~+ 2i \thb^2 \Fab \Tr (\Gbar_\a^{~\ad} \ast \Phib) \ast \Tr ( \partial_{\beta\ad} \Phi )
+ 2i \thb^2 \Fab \Tr ( \Gbar_\alpha^{~\ad} \ast \Phi) \ast \Tr ( \partial_{\beta\ad} \Phib ) \Big]
\Big\}
\non
\eea
and tune the renormalization of $\kappa$ with the
wave--function renormalization in order to make $SU(\Nc)$ and $U(1)$ superfields 
to renormalize in the same way. Consequently, a cubic superpotential of the form $h \int \Tr \Phi_\ast^3 
+ \bar{h} \int \Tr \Phib_\ast^3$ can be safely added, with no need of further terms like the ones 
in (\ref{generalized}). 

As discussed in details in Appendix A, the background field method can be easily generalized to the 
action (\ref{quadratickappa}) by performing a change of variables 
$\Phibold_q \to \Phibold_q' = (\Phibold_q^a, \kappa_1 \Phibold_q^0)$ and 
$\Phibbold_q \to \Phibbold_q' = (\Phibbold_q^a, \kappa_2 \Phibbold_q^0)$, $\kappa_1 \kappa_2=\kappa$, in the 
functional integral. The net result is a rescaling of the covariant propagators according to eqs. 
(\ref{rescaledprop1}-\ref{rescaledprop4}). Expanding the propagators in powers of the background gauge fields 
(see Appendix A) this is equivalent to a rescaling of the abelian propagator 
\beq
\langle \phib^0 \phi^0 \rangle = \frac{1}{\kappa} ~\frac{1}{\Box_0} 
\label{rescaling}
\eeq
and a rescaling of all gauge--chiral interaction vertices involving
abelian superfields. Precisely, vertices containing $\Phibold^0$, $\Phibbold^0$ acquire an extra coupling 
constant $1/\kappa_1$, $1/\kappa_2$, respectively. 

It is important to note that in the covariant propagators
the $\kappa_1, \kappa_2$ couplings appear only in terms proportional to the deformation parameter. 
Therefore, the dependence on these two couplings would disappear in the ordinary $N=1$ supersymmetric 
case. In that case, as it is well known,  the rescaling (\ref{rescaling}) of the abelian propagator would be the only effect of  
choosing a modified quadratic lagrangian for the abelian superfields. 

\vskip 15pt
To summarize, we begin with a NAC classical gauge theory whose gauge sector is still described by
(\ref{min1}) or (\ref{min2}) , whereas the matter action is given by (\ref{quadratickappa}) supplemented by
the single--trace cubic superpotential. However, as appears from one--loop calculations, extra couplings need be considered 
which are consistent with $N=1/2$ supersymmetry and supergauge invariance. In the 
next Section we will select all possible couplings which can be added at classical level.

\section{The most general gauge invariant action}

Before entering the study of renormalization properties, we will select all possible 
divergent structures which could come out at quantum level on the basis of dimensional analysis and 
global symmetries of the theory.

\subsection{Dimensional analysis and global symmetries}

The most general divergent term which may arise at quantum level has the form
\bea
\label{counterterm1}
  \intsup \thb^{\bar{\tau}}~ {\cal F}^\a~ \Lambda^\beta~
  D^\gamma~ \Db^{\bar{\gamma}}~ \partial^\delta~
  \Gbar^{\bar{\sigma}}~ \Phi^n~ \Phib^m~ h^r~ \bar h^s
\eea
where all the exponents are non negative integers. Of course, powers of the gauge coupling $g$ can
appear. However, its presence is irrelevant for our discussion, being $g$ adimensional and with
zero R--symmetry charge. Therefore, in what follows we will neglect it. 
 
We make the following simplifications:  
\begin{itemize}
  \item
  We can choose the connections
  to be the bosonic $\Gbar^{\a\ad}$. In fact, 
  thanks to the relation $\Gbar^{\a\ad} = -i D_\a \Gbar_\ad$, switching from bosonic 
  to fermionic connections would amount to shifting $\gamma \rightarrow \gamma + \bar\sigma$.

  \item 
  The parameter $\bar{\tau}$ takes the values $0,1,2$. However, we can fix it to be $2$
  by writing $\thb^\ad = \Db^\ad \thb^2 \rightarrow \thb^2 \Db^\ad$ and $-1 = \Db^2
    \thb^2 \rightarrow \thb^2 \Db^2$ where we think of integrating by parts the antichiral derivatives.
  \item 
  Assuming that the NAC deformation is a soft supersymmetry breaking mechanism we set $\beta=0$.
  \item 
  At one--loop, the $\Phi^3$ vertex provides a single power of the $h$ coupling 
  and one external $\Phi$-field. Taking into account that further external chirals can come 
  from gauge--chiral vertices, we have the constraint $r \leq n$. Similarly, for the antichiral
    vertex it must be $s \leq m$.
\end{itemize}

Therefore, the general structure for divergences can be reduced to the following form
\bea
\label{counterterm}
  \intsup \thb^2~ {\cal F}^\a~ \nabla^\gamma~
  \Db^{\bar\gamma} \partial^\delta~ \Gbar^{\bar{\sigma}}~ \Phi^n~
  \Phib^m~ h^r~ \bar h^s \qquad r \leq n \quad , \quad s \leq m
\eea
where the number of $\nabla$--derivatives should not exceed $(\bar{\sigma} + 2(n-1))$ in order
to avoid the integrand to be a total $\nabla$--derivative. 
Further constraints on the exponents come from imposing the global symmetries as listed in Table 1,
in addition to the request for the integrand to have  
mass dimension $2$. Moreover, we need impose the number of dotted and undotted indices to be 
even from the requirement that they contract among themselves to generate a supersymmetry
singlet. Finally, we impose $\alpha \geq 1$ to allow for 
a non--trivial dependence on the nonanticommutative parameter.

With the charge assignements given in Table 1 the set of constraints read
\beq
\begin{tabular}{ll}
Dimensions: & $-3 - \alpha + \frac{\gamma}{2} +\frac{\bar\gamma}{2}
+ \delta + \bar\sigma + n + m = 0$ 
\\ 
R-charge: & $2 - 2\alpha +
\gamma - \bar\gamma - n + m + r - s = 0$ 
\\ 
Index contraction: &
$2\alpha + \gamma + \delta + \bar\sigma = 2 l + 4$ 
\\ 
& $\bar\gamma +
\delta + \bar\sigma = 2 l'$ 
\\ 
Derivatives: & $\gamma \leq \bar\sigma + 2n - 2$ 
\\ 
$\Phi$--symmetry: & $n - m + 3( s - r ) = 0$ 
\\
One--loop rules: & $r \leq n$ \\ & $s \leq m$
\\ 
\label{system}
\end{tabular}
\eeq
where $l, l' \geq 0$ are integer numbers.

\begin{table}
\label{tab:dim-R}
\begin{center}
\begin{tabular}{|c|c|c|r|}
\hline
  &  dim & R-charge & $\Phi$-charge \\
\hline
$\overline{\Gamma}^{\a\ad}$ & $1$    &  $0$ & $0$ \\
\hline
$D_\alpha\equiv \nabla_\alpha$ &   $1/2$   & $1$ & $0$ \\
\hline
$\Db_\ad$ &   $1/2$   & $-1$ & $0$ \\
\hline
$\thb$ & $- 1/2$ & $1$ & $0$ \\
\hline
$\partial_{\a\ad}$ &  $1$   &  $0$ & $0$ \\
\hline
${\cal F}^{\rho \gamma}$ &   $-1$  &  $-2$ & $0$ \\
\hline
$\Phi$ & $1$ & $-1$ & $1$ \\
\hline
$h$ & $0$ & $1$ & $-3$ \\
\hline
$\Phib$ & $1$ & $1$ & $-1$ \\
\hline
$\barh$ & $0$ & $-1$ & $3$ \\
\hline
\end{tabular}
\caption{Dimensions, R and $\Phi$--charge assignments of $N=1/2$ operators.}
\end{center}
\end{table}

Combining the first two equations we get
\beq
  8 -4l' = 3n +m -r +s \geq 3n +m -r \geq 2n +m \geq 0
\eeq
from which we derive the conditions
\beq
\label{lprime}
l' \leq 2 \nonumber \qquad \quad 2n +m \leq 8 -4l'
\eeq
A simple constraint on $l$ can be obtained from merging the third, the forth and the sixth 
equations in (\ref{system})
\bea
  &&2(l-l')+4 = 2\alpha + \gamma - \bar\gamma \leq 2\alpha + \bar\sigma
  +2n - 2 -\bar\gamma 
  \nonumber \\ 
  &&~~~~~~~~~~~~~~~~ = 2\alpha + 2 l' -\delta +2n -2\bar\gamma -2
  \nonumber \\ 
  &&\Rightarrow l \leq \alpha +
  2l' -3 -\frac{1}{2} \delta +n -\bar\gamma
\eea
Then, using the first constraint and the previous bound we find
\beq
2n + m = 3 +\alpha +n -\frac{1}{2} \gamma -\frac{1}{2} \bar\gamma
-2l' + \bar\gamma \leq 8 -4l' 
\eeq  
which, after a bit of trivial algebra, provides a constraint on $\a$
\beq
\label{eq:alpha bound}
1 \leq \alpha \leq  4 -l' -\bar\gamma
\eeq
Finally, using this condition we can constrain $l$ even more and obtain
\beq
\label{eq:l bound}
0 \leq l \leq 5 -\delta -l' - 2\bar\gamma
\eeq

Now we are ready to list divergent contributions. We assign values $0,1,2$ to $l'$
according to (\ref{lprime}), 
and we fix $\delta$, $\bar\sigma$ and $\bar\gamma$, which are bounded
by $l'$ itself. Then we can vary $l$ into the range given by
(\ref{eq:l bound}) and $\alpha$ in the range (\ref{eq:alpha bound}),
while the value of $\gamma$ follows immediately from the third
equation in (\ref{system}). Finally, the remaining parameters
($n,m,r,s$) are varied according with the set of equations
(\ref{system}).

A detailed investigation reveals that, independently of their particular trace structure,
the only allowed terms are (for the moment we forget about $\ast$--products) 

\begin{enumerate}
\item 
Matter sector. These structures are obtained by setting $\bar{\sigma}=0$ when $l'=0,1$
and correspond to 
\bea
&\bullet&  \barh (h \barh)^r {\cal F}^2 \intsup \thb^2 \Phi \Phib^4 \qquad \qquad r = 0,1
\label{c1}
\\
&\bullet&   (h \barh)^r {\cal F}^2 \intsup \thb^2 \Phi (\nabla^2 \Phi)   \Phib^2 \qquad  r \leq 2
\label{c2}
\\
&\bullet& h ~{\cal F}^2 \intsup \thb^2 \Phi (\nabla^2 \Phi)^2 
\label{c3}
\\
&\bullet&  h ~\Fab \intsup \thb^2 (\nabla_\a \Phi) (\nabla_\b \Phi)  \Phi
\label{c4}
\eea
Powers of the gauge coupling $g$ are also allowed. 
The first three terms are non--vanishing whatever 
the color structure is. In the abelian case they correspond to the actual structures 
which arise at one and two loops in the ungauged NAC WZ model \cite{GPR1, BFR, JJP}.
The last term, instead, is nontrivial only  when  $\nabla_\a \Phi$ and $\nabla_\b \Phi$ have different color index.
Therefore, it is present only when gauging the WZ model with a non--abelian group.

\item
Mixed sector. All structures selected correspond to the case $l'=1$ and are
given by
\bea
&\bullet&  (h \barh)^r \Fab \intsup \thb^2
\partial_{\beta}^{~\ad} \Gbar_{\alpha\ad} \Phi
\Phib 
\label{gc1}
\\
&\bullet& (h \barh)^r \Fab \intsup \thb^2
\Gbar_{\beta}^{~\ad} \Gbar_{\alpha\ad} \Phi \Phib
\label{gc2}
\\
&\bullet& \barh ~{\cal F}^2 \intsup \thb^2
\Gbar^{\alpha\ad} \Gbar_{\alpha\ad} \Phib \Phib
\Phib 
\label{gc3}
\\ 
&\bullet& (h \barh)^r {\cal F}^2 \intsup \thb^2
\Wbar^\ad \Wbar_\ad \Phi \Phib 
\label{gc4}
\eea
where in (\ref{gc1}) the space--time derivative can act on any of the three fields. 

At one--loop, we can only have $r = 0,1$. When $r=0$ a $g^2$ factor is present and corresponds 
to contributions generated by mixed gauge--chiral vertices. When $r=1$ we have divergent terms 
generated by pure (anti)chiral vertices.

\item
Gauge sector. This case corresponds to $l' = 2$ because of the bound 
$2n+m \leq 8-4l'=0$ which implies $n=m=0$, i.e. no external (anti)chiral fields.
The structures we find are exactly the ones found in \cite{GPR2}. 
    
\end{enumerate}

The previous analysis can be generalized to the case $\b \neq 0$ in (\ref{counterterm1}) allowing for positive powers of the 
UV cut--off. It is not difficult to see that for any positive value of $\b$ non--trivial structures which satisfy all the constraints cannot be constructed. This proves that even in the presence of interacting matter
supersymmetry is softly broken.

\subsection{Gauge invariance}

The previous structures have been selected without requiring supergauge invariance. We expect that
imposing it as a further constraint, only particular linear combinations of the previous terms
with specific color structures will survive. 

In the matter sector,   thanks to the presence of the $\thb^2$ factor, the (anti)chiral interaction terms 
(\ref{c1}--\ref{c3}) are gauge--invariant, independently of their color structure.
The term (\ref{c4}) is non--vanishing only when it is single--trace and it is gauge invariant.

Focusing on the mixed sector, it is easy to see that the general terms (\ref{gc3}, \ref{gc4}) 
are always gauge invariant, independently of their trace structure.

Terms (\ref{gc1}, \ref{gc2}), instead, give rise to different gauge invariant 
combinations depending on their trace structure.
The only invariant single--trace operator which can arise at one--loop is
\beq
\Fab \thb^2 \Tr \left( \partial_{\beta\ad}
\Gbar_\alpha^{~\ad} \{ \Phi, \Phib \} - \frac{i}{2}
[\Gbar_{\beta\ad}, \Gbar_\alpha^{~\ad} ]_\ast \{ \Phi, \Phib \} \right)
\eeq
where the explicitly indicated $\ast$-product is the only non--trivial $\ast$--product 
which appears.
Looking at double--trace operators, we already know that structures of the
form (\ref{gc1}, \ref{gc2}) combine with the double--trace 2pt function in order to 
make it gauge invariant (see eqs. (\ref{gauge1}, \ref{gauge2})). Further gauge 
invariant combinations from (\ref{gc1}, \ref{gc2}) are
\bea
  &&\Fab \thb^2~ \Tr \left( \partial_{\beta\ad}
  \Gbar_\alpha^{~\ad} \Phi - \frac{i}{2} [\Gbar_{\beta\ad},
    \Gbar_\alpha^{~\ad} ]_{\ast} \Phi \right) \Tr ( \Phib) 
  \label{eqn:gauge2} 
  \\ 
  &&\Fab \thb^2~ \Tr \left( \partial_{\beta\ad}
  \Gbar_\alpha^{~\ad} \Phib - \frac{i}{2} [\Gbar_{\beta\ad},
    \Gbar_\alpha^{~\ad} ]_{\ast} \Phib \right) \Tr ( \Phi) 
  \label{eqn:gauge4} 
  \\ 
  &&\Fab \thb^2~ \Tr \left(
  \partial_{\beta\ad} \Gbar_\alpha^{~\ad} \right) \Tr \left( \Phi
  \Phib \right) 
  \label{eqn:gauge3}
\eea
while there is no way to saturate the gauge variation of $\,\Fab
\thb^2 \Tr \left( \Gbar_\alpha^{~\ad} \right) \Tr \left(
(\partial_{\beta\ad} \Phi) \Phib \right)$ or, similarly, of the term
obtained by exchanging $\Phi \leftrightarrow \Phib$. 
Indeed, only the combination 
$\left[\Fab \thb^2 \Tr \left( \Gbar_\alpha^{~\ad} \right) \Tr \left(
(\partial_{\beta\ad} \Phi) \Phib \right) + 
\Fab \thb^2 \Tr \left( \Gbar_\alpha^{~\ad} \right) \Tr \left(
\Phi (\partial_{\beta\ad} \Phib)  \right) \right]$ is gauge invariant. However, integrating
by parts, this reduces to (\ref{eqn:gauge3}).

Using similar arguments, we find that the only triple--trace gauge--invariant
operator is
\beq
\Fab \thb^2~ \Tr \left( \partial_{\beta\ad}
  \Gbar_\alpha^{~\ad} \right) \Tr ( \Phi ) \Tr ( \Phib)
\eeq

Finally, looking at the gauge sector, once we impose gauge invariance only terms corresponding
to all NAC structures present in (\ref{min1}, \ref{min2}) are selected.

\subsection{The general action}

We are now ready to propose the most general classical action for a NAC gauge theory with massless
matter in the adjoint of $SU({\cal N}) \otimes U(1)$. Introducing the greatest number of coupling
constants compatible with gauge invariance, we write
\beq
S = S_{gauge} + S_{matter} + S_{\Gbar} + S_{\Bar{W}}
\label{main}
\eeq 
where $S_{gauge}$ is given in (\ref{min1}) (or equivalently (\ref{min2})), 
\bea
\label{mainmatter}
S_{matter} &=&  \intsup  \Big\{ \Tr \left( \Phib \ast \Phi \right)
+ \frac{\kappa - 1}{\Nc} \Big[ \Tr \Phib  \ast \Tr \Phi 
\non \\
&&~~~\quad ~~~+ 2i \thb^2 \Fab \Tr (\Gbar_\a^{~\ad} \ast \Phib) \ast \Tr ( \partial_{\beta\ad} \Phi )
+ 2i \thb^2 \Fab \Tr ( \Gbar_\alpha^{~\ad} \ast \Phi) \ast \Tr ( \partial_{\beta\ad} \Phib ) \Big]
\Big\}
\non \\
&+& h \intch \Tr \Phi_\ast^3 + \bar{h} \intach \Tr \Phib_\ast^3
+\tilde{h}_3 ~\Fab \intsup \thb^2 \Tr( (\nabla_\a \Phi)(\nabla_\b \Phi) \Phi)
\non\\
&+&  \sum_{j=1}^3 \, h_3^{(j)} \, {\cal C}_j^{ABC} \, 
{\cal F}^2 \intsup \thb^2 \Phi^A (\nabla^2 \Phi^B) (\nabla^2 \Phi^C)
\non \\
&+&  \sum_{j=1}^{10} \, h_4^{(j)} \, {\cal D}_j^{ABCD} \, {\cal F}^2 \intsup \thb^2 \Phi^A
(\nabla^2 \Phi^B) \Phib^C \Phib^D
\non
\\
&+& \sum_{j=1}^{12} \, h_5^{(j)} \, {\cal E}_j^{ABCDE} \, {\cal F}^2 \intsup \thb^2 \Phi^A \Phib^B
\Phib^C \Phib^D \Phib^E 
\eea
and $S_{\Gbar}, S_{\Wb}$ contain all possible gauge invariant mixed terms 
proportional to the bosonic connection 
\bea
S_{\Gbar} &=& t_1 ~\Fab \intsup \thb^2 \Tr \left(
  \partial_{\beta\ad} \Gbar_\alpha^{~\ad} \right)\; \Tr \left(
  \Phib \Phi \right) 
  \nonumber \\ 
  &+& t_2 ~\Fab \intsup \thb^2 \Tr
  \left( \partial_{\beta\ad} \Gbar_\alpha^{~\ad} \right)\; \Tr
  \Phib\; \Tr \Phi 
  \nonumber \\ 
  &+& t_3 ~\Fab \intsup \thb^2 \Tr
  \left( ( \partial_{\beta\ad} \Gbar_\alpha^{~\ad} - \frac{i}{2} [
    \Gbar_{\beta\ad}, \Gbar_\alpha^{~\ad} ]_\ast ) \; \{ \Phib,
  \Phi \} \right) 
  \nonumber \\ 
  &+& t_4 ~\Fab \intsup \thb^2 \Tr
  \left( ( \partial_{\beta\ad} \Gbar_\alpha^{~\ad} - \frac{i}{2} [
    \Gbar_{\beta\ad}, \Gbar_\alpha^{~\ad} ]_\ast )\; \Phi
  \right)\; \Tr \Phib 
  \nonumber \\ 
  &+& t_5~\Fab \intsup \thb^2
  \Tr \left( ( \partial_{\beta\ad} \Gbar_\alpha^{~\ad} -
  \frac{i}{2} [\Gbar_{\beta\ad}, \Gbar_\alpha^{~\ad} ]_\ast )\;
  \Phib \right)\; \Tr \Phi \non \\
  &+& \sum_{j=1}^{18} \, \tilde t_6^{(j)} \, {\cal G}_j^{ABCDE} 
  \, {\cal F}^2 \intsup \thb^2 \, \Gbar^{A \, \a\ad}
 \Gbar_{\a\ad}^B \Phib^C \Phib^D \Phib^E 
  \label{Gammabar}
\eea
and to the field--strength
\beq
\label{Wbar}
S_{\Wbar} =  \sum_{j=1}^{12} \, l_j \, {\cal H}_j^{ABCD} \, {\cal F}^2 \intsup \thb^2~ 
 \Wbar^{A \, \ad} \Wbar_\ad^B  \Phi^C \Phib^D 
\eeq  
We have introduced the following group tensors to take into account all possible color structures
(we use the shorten notation $\Tr(T^A) = (A)$ for any group matrix) 
\bea
&& {\cal C}^{ABC}_1 = (ABC) \qquad {\cal C}^{ABC}_2 = (AB)(C) \qquad {\cal C}^{ABC}_3 = (A)(B)(C)
\non \\
~~\non \\
&& {\cal D}^{ABCD}_1 = (ABCD) \quad {\cal D}^{ABCD}_2 = (ACBD) 
\non \\
&& {\cal D}^{ABCD}_3 = (A)(BCD) \quad {\cal D}^{ABCD}_4 = (C)(ABD)
\non \\
&& {\cal D}^{ABCD}_5 = (AB)(CD) 
\quad {\cal D}^{ABCD}_6 = (AC)(BD) 
\non \\
&& {\cal D}^{ABCD}_7 = (AB)(C)(D)
\quad {\cal D}^{ABCD}_8 = (AC)(B)(D) \quad {\cal D}^{ABCD}_9 = (A)(B)(CD) 
\non \\
&&{\cal D}^{ABCD}_{10} = (A)(B)(C)(D)
\non \\
&& ~~\non \\
&& {\cal E}^{ABCDE}_1 = (ABCDE) \quad {\cal E}^{ABCDE}_2 = (ABCD)(E) 
\quad {\cal E}^{ABCDE}_3 = (BCDE)(A)
\non \\
&& {\cal E}^{ABCDE}_4 = (ABC)(DE) 
\quad {\cal E}^{ABCDE}_5 = (BCD)(AE) \quad {\cal E}^{ABCDE}_6 = (ABC)(D)(E)
\non \\
&& {\cal E}^{ABCDE}_7 = (BCD)(A)(E) \quad {\cal E}^{ABCDE}_8 = (AB)(CD)(E) 
\quad {\cal E}^{ABCDE}_9 = (BC)(DE)(A)
\non \\
&& {\cal E}^{ABCDE}_{10} = (A)(BC)(D)(E) \quad {\cal E}^{ABCDE}_{11} = (AB)(C)(D)(E) 
\non \\
&&{\cal E}^{ABCDE}_{12} = (A)(B)(C)(D)(E)
\non \\
&& ~~\non \\
&& {\cal G}^{ABCDE}_1 = (ABCDE) \quad  {\cal G}^{ABCDE}_2 = (ACBDE) 
\non \\
&&{\cal G}^{ABCDE}_3 = (ABCD)(E)  \quad {\cal G}^{ABCDE}_4 = (ACBD)(E) 
\quad {\cal G}^{ABCDE}_5 = (BCDE)(A)
\non \\
&& {\cal G}^{ABCDE}_6 = (ABC)(DE) 
\quad {\cal G}^{ABCDE}_7 = (BCD)(AE) \quad {\cal G}^{ABCDE}_8 = (AB)(CDE) 
\non \\
&&{\cal G}^{ABCDE}_9 = (ABC)(D)(E)
\quad {\cal G}^{ABCDE}_{10} = (BCD)(A)(E) \quad {\cal G}^{ABCDE}_{11} = (A)(B)(CDE)
\non \\
&&{\cal G}^{ABCDE}_{12} = (AB)(CD)(E) 
\quad {\cal G}^{ABCDE}_{13} = (BC)(DE)(A) \quad  {\cal G}^{ABCDE}_{14} = (BC)(AD)(E)
\non \\
&& {\cal G}^{ABCDE}_{15} = (A)(BC)(D)(E) \quad {\cal G}^{ABCDE}_{16} = (AB)(C)(D)(E)
\non \\
&&{\cal G}^{ABCDE}_{17} = (A)(B)(CD)(E)
\quad {\cal G}^{ABCDE}_{18} = (A)(B)(C)(D)(E)
\non \\
&& ~~\non \\
&& {\cal H}^{ABCD}_1 = (ABCD) \quad {\cal H}^{ABCD}_2 = (ACBD) 
\non \\
&&{\cal H}^{ABCD}_3 = (A)(BCD) \quad {\cal H}^{ABCD}_4 = (C)(ABD)
\quad {\cal H}^{ABCD}_5 = (D)(ABC) 
\non \\
&&{\cal H}^{ABCD}_6 = (AB)(CD) \quad {\cal H}^{ABCD}_7 = (AC)(BD)
\non \\
&& {\cal H}^{ABCD}_8 = (AB)(C)(D) \quad {\cal H}^{ABCD}_9 = (AC)(B)(D) 
\quad {\cal H}^{ABCD}_{10} = (AD)(B)(C)
\non \\
&& {\cal H}^{ABCD}_{11} = (A)(B)(CD) \quad {\cal H}^{ABCD}_{12} = (A)(B)(C)(D) 
\eea 
Whenever in the action the $\ast$--product is not explicitly indicated the products are 
indeed ordinary products. This happens in most terms above because of the presence of 
the $\thb^2$ factor.

\section{One--loop renormalizability and gauge invariance}

In this Section we will provide general arguments in support of the one--loop renormalizability 
of the action (\ref{main}).

The action (\ref{main}) has been obtained by including all possible divergent structures which can appear at 
one--loop. Therefore, one might be tempted to conclude that it is {\em a fortiori} renormalizable. However, 
some of these terms need enter particular linear combinations in order to insure
gauge--invariance. Such terms are identified by couplings $(\kappa -1)$ and $t_{3}, t_4, t_5$. 
Therefore, proving one--loop renormalizability amounts to  prove that quantum corrections maintain the 
correct gauge--invariant combinations.
In what follows we will be mainly focused on these terms and find the conditions under which gauge 
invariance is maintained at quantum level. 

In order to perform one--loop calculations we use the background--field method revised in Section 2 and 
applied to the general action (\ref{main}). In Appendix A the necessary
Feynman rules are collected.

When drawing possible divergent diagrams we make use of the following observations: 
First of all, from the dimensional analysis performed in Section 5, one--loop 
divergences  may be  proportional to the non--anticommutation parameter ${\cal F}$ at most quadratically. 
Therefore, we do not take into account diagrams which give higher powers of ${\cal F}$. 
Moreover, the structures we are mainly interested in (the ones associated to the couplings 
$(\kappa -1)$ and $t_{3},t_4,t_5$) are proportional to $\Fab$, so they cannot receive corrections 
from diagrams which contain vertices proportional to ${\cal F}^2$. 

For each supergraph we perform $\Del$--algebra \cite{superspace, GSZ,
GZ} in order to reduce it to an ordinary momentum graph and read the
background structures associated to the divergent integrals.  We
discuss renormalizability of the different sectors, separately.

\subsection{Pure gauge sector}

In the absence of a superpotential term, the one--loop effective action for
the gauge sector has been already computed in \cite{GPR2}. 

With the addition of the cubic superpotential and the related modifications of the classical action, 
the gauge effective action could, \emph{a priori}, get corrected because of
two different reasons: The modification of the chiral propagators to include 
different couplings for the abelian superfields which might affect the evaluation of $\Delta$ in (\ref{effective}), 
and the presence of new mixed gauge--chiral interaction vertices from $S_{int}$ in (\ref{effective0}) as
coming from $S_{\Gbar}$ and $S_{\Wb}$ and the second line of (\ref{mainmatter}).

The former modification is harmless  because of the reparametrization invariance of 
$\Delta$ under the change of variables
$\Phibold^A \rightarrow \Phibold'^A \equiv
(\Phibold^a, \kappa_1 \Phibold^0)$,
$\Phibbold^A \rightarrow \Phibbold'^A \equiv
(\Phibbold^a, \kappa_2 \Phibbold^0)$, $\kappa = \kappa_1
\kappa_2$ 
\bea
  \Delta &=&\int D \Phibold D \Phibbold~ \exp{ \intsup \left( \Tr \Phibbold
    \Phibold + \frac{\kappa -1}{\mathN} \Tr \Phibbold \Tr \Phibold \right) }
  \non
  \\ 
  &&\sim \int D \Phibold' D \Phibbold'~ \exp{ \intsup \Tr \Phibbold' \Phibold' }
\eea
The $\kappa$--independence of $\Delta$ can be also checked by explicit calculations, noting that 
in its one--loop expansion abelian superfields never enter. 

The other source of possible modifications for the gauge effective action 
is the appearance of new gauge--chiral vertices in $S_{\Gbar}$ 
and $S_{\Wb}$, eqs. (\ref{Gammabar}) and (\ref{Wbar}), and second line of (\ref{mainmatter}). 
In any case the new vertices produce tadpole--like diagrams when contracting 
the matter superfields leaving gauge fields as background fields. After $\Del$--algebra,
the tadpole provides the covariant propagator $1/\Box_{cov}$ which can be expanded as 
in (\ref{eqn:box_{cov}}) up to second order in $\Gbar$ producing divergent contributions.  
It is easy to prove that these divergences cancel exactly as in the ordinary case.

We conclude that the addition of a cubic superpotential and related modifications does
not change the results in \cite{GPR2} for the divergent part of the one--loop gauge effective action. 
Therefore, if we start with a classical action as the one in (\ref{min1}) or (\ref{min2}) 
we can multiplicatively renormalize all the divergences of the gauge sector (see Ref. \cite{GPR2}
for the detailed calculation).

\subsection{Gauge--matter sector} \label{sec:gauge--matter}

We now study one--loop divergent contributions to the rest of the
action, i.e.  $S_{matter} + S_{\Gbar} + S_{\Wb}$ (see
eqs. (\ref{mainmatter}--\ref{Wbar})).  The contributions identified by
the couplings $(\kappa -1)$ and $t_3,t_4,t_5$, whose gauge invariance
is under discussion, belong to this sector. Therefore, we concentrate primarily on this kind of terms.

\begin{figure}
  \includegraphics[width=\textwidth]{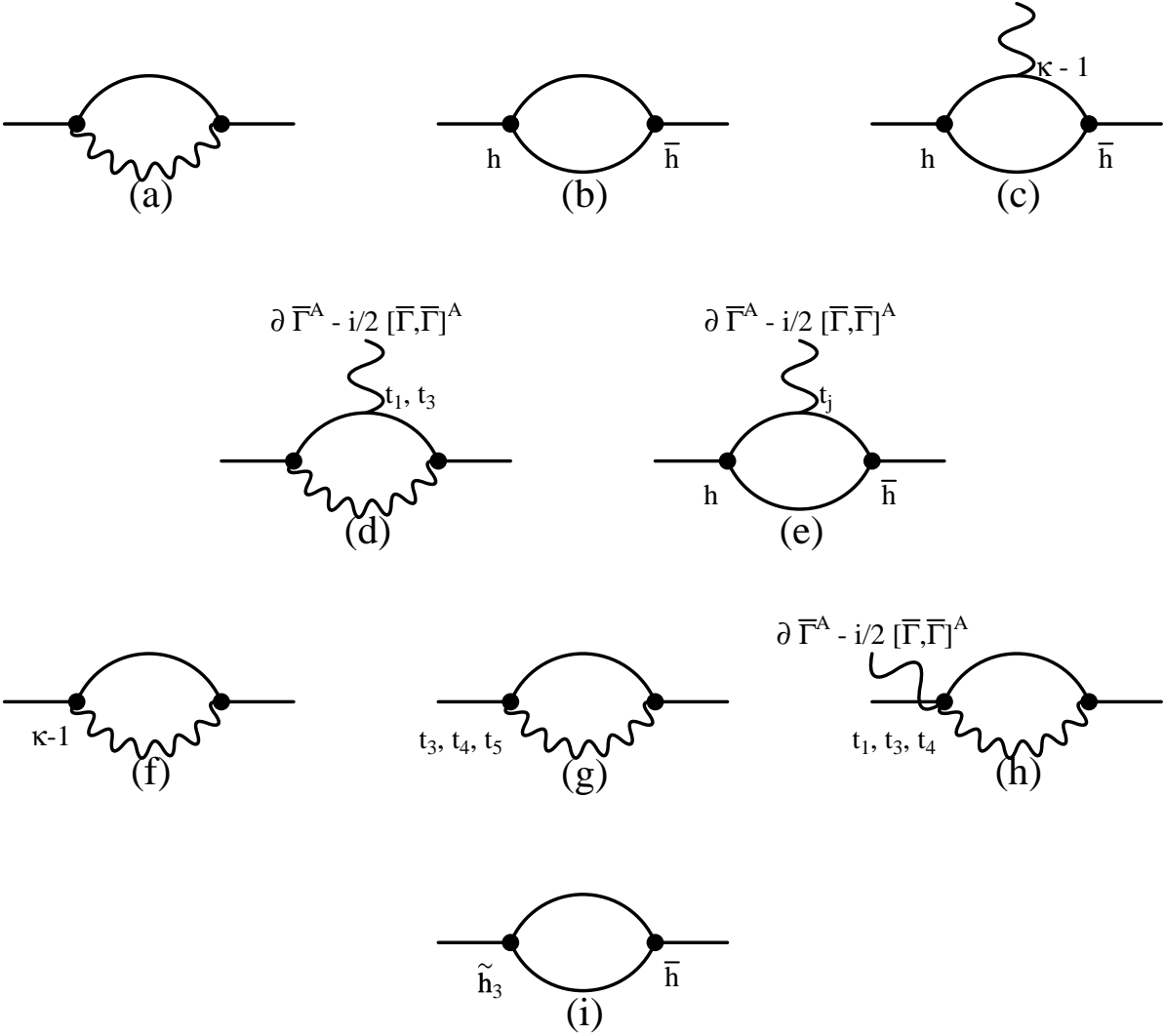}
  \caption{Master diagrams which, after the expansion of the covariant propagators, give rise to 
  two, three and four--point divergent contributions.}
  \label{fig:diagrams}
\end{figure}

Divergent contributions come from diagrams in Fig. 2 where internal lines are covariant gauge
and chiral propagators (see eqs. (\ref{vectorprop}, \ref{rescaledprop1}--\ref{rescaledprop4})) . 
Expanding the propagators in powers of the background superfields 
we find two, three and four--point divergences, whereas higher powers give rise to finite
contributions. 

We analyze the diagrams separately. 

\vskip 10pt
\noindent
\underline{Diagram (2a)} 

\noindent
Diagram (2a) is obtained by joining two vertices in
Fig. (\ref{fig:vertices}a) by one chiral propagator $1/\Box_-$ and one
vector propagator $1/\hat{\Box}$. Expanding the propagators at the
lowest order, $1/\Box_-, 1/\hat{\Box} \sim 1/\Box$, we obtain the
ordinary divergent quadratic term when the $\ast$--product at the
vertices is neglected.  Quadratic terms with a nontrivial dependence
on ${\cal F}$ are finite.  Instead, divergent three--point functions
exhibiting a linear dependence on ${\cal F}$ come from the first order 
expansion of the propagators (see eqs.(\ref{eqn:box_{vector}}, \ref{eqn:box_{cov}})).
Their dependence on the NAC parameter comes either when expanding
the $\ast$--product at the vertices or from $\Fab$ terms in 
eqs. (\ref{eqn:box_{vector}}, \ref{eqn:box_{cov}}). 
Combining all contributions, diagram (2a) gives rise to
\beq
\G^{(1)}_2(g)
+ \G^{(1)}_3(g) + \G'^{(1)}_{3}(g) + \G^{(1)}_4(g)
\eeq
where
\bea
\label{diver1}
&&  \Gamma^{(1)}_2(g) + \Gamma^{(1)}_3(g) = 2g^2 \mathcal{S} \intsup \Big[
  - \mathN \Tr \left( \Phibbold \Phibold \right)  
 \\
&& +
\Tr \bold{\Phib}  \ast \Tr \bold{\Phi} + 2i \thb^2 \Fab~ \Tr
  ( \Gbar_{\a}^{~ \ad} \ast \Phibbold ) \ast \Tr (
  \partial_{\beta \ad} \Phibold ) 
  +2i \thb^2 \Fab~ \Tr
  ( \Gbar_{\a}^{~ \ad} \ast \Phibold ) \ast \Tr (
  \partial_{\beta \ad} \Phibbold )
  \Big]
  \non
\eea
and
\beq
\G'^{(1)}_{3}(g) = 4ig^2 {\cal S} \Fab \intsup \thb^2 \Tr (
  \partial_{\beta\ad} \Gbar_\alpha^{~\ad} ) \Tr ( \Phibold \Phibbold )
\label{diver2}
\eeq
Four--point functions $\G^{(1)}_{4}(g)$ come from the second order expansion of the 
product of the two propagators. They are divergent but always proportional to ${\cal F}^2$, therefore
automatically gauge--invariant. 
  
We note that the divergences (\ref{diver1}, \ref{diver2}) come in the right linear combinations
for preserving gauge--invariance.

\vskip 10pt
\noindent
\underline{Diagrams (2b, 2c)}
 
\noindent 
With the aim of discussing gauge invariance, it is convenient to
consider the sum of diagrams (2b) and (2c). Diagram (2b) is obtained by
joining one $h$ and one $\bar{h}$ vertices in
Fig. (\ref{fig:vertices}h, \ref{fig:vertices}j) by two $1/\Box_-$ propagators, whereas
diagram (2c) is generated from diagram (2b) by the insertion of an
extra $(\kappa-1)$--vertex in Figs. (3c, 3d).  Expanding
the chiral propagators at the lowest order, $1/\Box_- \sim 1/\Box$
and neglecting the $\ast$--product at the vertices, from diagram (2b)
we obtain the ordinary divergent quadratic term and from diagram (2c) a
three--point divergent contribution linear in ${\cal F}$.  Further
three and four--point contributions come from the higher order
expansion of the propagators in both diagrams.  In diagram
(2b) the linear dependence in the NAC parameter comes either from terms
in the propagator expansion or from the $\ast$--product at the
vertices.

Combining all contributions, the sum of the two diagrams gives rise to 
\beq
\G^{(1)}_2(h,\barh)
+ \G^{(1)}_3(h,\barh) + \G'^{(1)}_{3}(h,\barh) + \G^{(1)}_4(h,\barh)
\eeq
where
\bea
&&\G^{(1)}_2(h,\barh) + \G^{(1)}_3(h,\barh) = 
\mathcal{S} \intsup \Big\{ 9h\overline{h}\left(\mathN + 4 \frac{1- \kappa}{\mathN\kappa} \right)
\Tr \left( \bold{\Phib} \ast \bold{\Phi} \right) 
\non \\ 
&+& 9h\overline{h} \left( 1 + 2 \left( \frac{1-\kappa}{\mathN\kappa} \right)^2 \right) 
\left[ \Tr \bold{\Phib}  \ast \Tr \bold{\Phi} + 2i \thb^2 \Fab~ \Tr
  ( \Gbar_{\a}^{~ \ad} \ast \Phibbold ) \ast \Tr (
  \partial_{\beta \ad} \Phibold ) \right. 
\non \\
&~&~~~~~~~~~~~~~~ \qquad \qquad \qquad
\left.   + 2i \thb^2 \Fab~ \Tr ( \Gbar_{\alpha}^{~ \ad} \ast \Phibold) 
\ast \Tr \left( \partial_{\beta \ad} \Phibbold \right) \right] \Big\}
\label{2pointmodified}
\eea
\bea
\label{eqn:3-point}
     \Gamma'^{(1)}_3 (h, \overline h) &= &\left[ \frac{54}{\mathN} -
      \frac{18}{\kappa\mathN} - \frac{18}{\kappa_1\mathN} -
      \frac{18}{\kappa_2\mathN} - 36\frac{1-\kappa}{\kappa\mathN}
      \right] i 
      \non \\ 
      &~&~~~~~~~~~~\times h\overline{h} \mathcal{S} \Fab
    \intsup \thb^2 \Tr ( \partial_{\beta\ad}
    \Gbar_\alpha^{~\ad} \{ \Phibold, \Phibbold \} ) 
    \non \\ 
    &+&\left[ -
      \frac{36}{\mathN^2} + \frac{36}{\kappa_1\mathN^2} +
      \frac{36}{\kappa\mathN^2} - \frac{36}{\kappa_1\kappa\mathN^2}
      \right] i 
      \non \\ 
      &~&~~~~~~~~~~\times h\overline{h} \mathcal{S} \Fab
    \intsup \thb^2 \Tr ( \partial_{\beta\ad}
    \Gbar_\alpha^{~\ad} \Phibbold )\; \Tr \Phibold 
    \non \\ 
    &+&\left[ - \frac{36}{\mathN^2} + \frac{36}{\kappa_2\mathN^2} +
      \frac{36}{\kappa\mathN^2} - \frac{36}{\kappa_2\kappa\mathN^2} -
      36 \left( \frac{1-\kappa}{\kappa\mathN} \right)^2 \right] i
    \non \\
     &&~~~~~~~~~~~\times h\overline{h} \mathcal{S} \Fab \intsup
    \thb^2 \Tr ( \partial_{\beta\ad} \Gbar_\alpha^{~\ad}
    \Phibold )\; \Tr \Phibbold  
    \non \\ 
    &+&\left[ -
      \frac{36}{\mathN^2} + \frac{36}{\kappa_1\mathN^2} +
      \frac{36}{\kappa_2\mathN^2} - \frac{36}{\kappa\mathN^2} - 72
      \left( \frac{1-\kappa}{\kappa\mathN} \right)^2 \right] i
    \non \\ 
    &~&~~~~~~~~~~\times h\overline{h} \mathcal{S} \Fab \intsup
    \thb^2 \Tr ( \partial_{\beta\ad} \Gbar_\alpha^{~\ad})\; 
    \Tr ( \Phibold \Phibbold ) 
    \non \\ 
    &+&\left[
      \frac{36}{\mathN^3} \frac{1-\kappa}{\kappa} \left( - 1 +
      \frac{1}{\kappa_1} + \frac{1}{\kappa_2} - \frac{1}{\kappa}
      \right) - 72 \left( \frac{1-\kappa}{\kappa\mathN} \right)^3
      \right] i 
     \non \\ 
      &~&~~~~~~~~~~\times h\overline{h} \mathcal{S} \Fab
    \intsup \thb^2 \Tr( \partial_{\beta\ad}
    \Gbar_\alpha^{~\ad} )\; \Tr \Phibold \; \Tr
    \Phibbold
 \eea
and 
\bea
\label{eqn:4-point}
  \Gamma^{(1)}_4 (h,\overline h)& = &-36 \left(
  \frac{1-\kappa}{\kappa\mathN} \right) h\overline h {\cal S} \Fab
  \intsup \thb^2 \Tr ( [\Gbar_{\beta\ad},
    \Gbar_\alpha^{~\ad}] \{ \Phibold, \Phibbold \} ) 
    \nonumber \\ 
    &~& -36 \left( \frac{1-\kappa}{\kappa\mathN} \right)^2 h\overline{h}
  \mathcal{S} \Fab \intsup \thb^2 \Tr ( [\Gbar_{\beta\ad},
    \Gbar_\alpha^{~\ad}] \Phibold )\; \Tr  \Phibbold 
\eea
We note that $\G^{(1)}_2(h, \barh) + \G^{(1)}_3(h, \barh)$ gives a gauge--invariant 
correction to the quadratic action. 
On the other hand, in $\G'^{(1)}_3(h, \barh)$ the first three lines are not gauge invariant. Possible
gauge completions for these terms are contained in $\G^{(1)}_4(h, \barh)$ if the corresponding 
factors satisfy the following constraints
\bea
  &-& \frac{i}{2} \left[ \frac{54}{\mathN} - \frac{18}{\kappa\mathN} -
    \frac{18}{\kappa_1\mathN} - \frac{18}{\kappa_2\mathN} -
    36\frac{1-\kappa}{\kappa\mathN} \right] i = -36 \left(
  \frac{1-\kappa}{\kappa\mathN} \right) 
  \\ 
  &-& \frac{i}{2} \left[ -
    \frac{36}{\mathN^2} + \frac{36}{\kappa_1\mathN^2} +
    \frac{36}{\kappa\mathN^2} - \frac{36}{\kappa_1\kappa\mathN^2}
    \right] i = 0 
    \\ 
    &-& \frac{i}{2} \left[ - \frac{36}{\mathN^2} +
    \frac{36}{\kappa_2\mathN^2} + \frac{36}{\kappa\mathN^2} -
    \frac{36}{\kappa_2\kappa\mathN^2} - 36 \left(
    \frac{1-\kappa}{\kappa\mathN} \right)^2 \right] i = -36 \left(
  \frac{1-\kappa}{\kappa\mathN} \right)^2
\eea
Having introduced two independent couplings $\kappa_1, \kappa_2$ we have the freedom to fix  them 
in order to satisfy this set of equations. It is easy to see that a non--trivial solution is given by
\beq
  \kappa_1 = 1 \qquad , \qquad \kappa_2 = \kappa
  \label{kappa}
\eeq
with no further requests on $\kappa$. 
Therefore, these conditions provide the right prescription for computing (2b,2c)--type contributions to
the effective action while preserving background gauge invariance.

Given the solution (\ref{kappa}) and recalling eq. (\ref{rescaling2})
we conclude that the extra coupling in front of the abelian
quadratic action origins entirely from a rescaling of the antichiral
superfields.

\vskip 10pt
\noindent
\underline{Diagrams (2d)}

\noindent
Diagrams of type (2d) are obtained by inserting 
in diagram (2a) one $t_1$ or one $t_3$ vertex (the insertion of $t_2, t_4, t_5$ vertices would give 
diagrams with vanishing color factors). Expanding the propagators and 
considering only divergent terms linear in the deformation parameter, it is easy to see that the diagram
with the insertion of one $t_1$ vertex  gives divergent contributions of the form $t_1, t_2$  
in $S_{\Gbar}$, whereas the diagram with one $t_3$ vertex contributes to the $t_1, t_3, t_4, t_5$ structures.  
They all come out automatically in the right gauge--invariant combinations.

\vskip 10pt
\noindent
\underline{Diagrams (2e)}

\noindent
Diagrams of type (2e) are obtained by inserting 
in diagram (2b) one of the $t_j$ vertices. Expanding the propagators and 
considering only divergent terms linear in the deformation parameter, from diagrams with $t_1,t_2$ 
vertices gauge--invariant structures associated to $t_1$ and $t_2$ in $S_{\Gbar}$ arise.
From diagrams with the insertion of vertices $t_3, t_4, t_5$ the background structure proportional to 
$\pa_\b^{\, \ad} \Gbar_{\a\ad}$ combines with the structure $[ \Gbar_\b^{\, \ad}, \Gbar_{\a\ad}]$
to give gauge--invariant divergent contributions of the form $t_1, \cdots, t_5$.  

\vskip 10pt
\noindent
\underline{Diagrams (2f)}

\noindent
This kind of diagrams are obtained by contracting the $(\kappa -1)$ vertices with a
quantum gauge $V$--field (see Figs. (3e, 3f, 3g)) with the ordinary vertex in
Fig. (\ref{fig:vertices}a). Expanding the covariant propagators it is
easy to see that they are either vanishing or finite.

\vskip 10pt
\noindent
\underline{Diagrams (2g)}

This class of diagrams is constructed by contracting a $t_3, t_4, t_5$--vertex in Fig. (\ref{fig:vertices}p)
with the ordinary vertex  (\ref{fig:vertices}a) (diagrams with $t_1$ and $t_2$  vertices vanish for color reasons). 
Explicit calculations reveal that nontrivial cancellations occur, so that no divergent 
contributions arise proportional to $t_4$ and $t_5$, whereas a non--vanishing term is generated 
by $t_3$ which is automatically in the right linear combination to respect gauge invariance. 
Precisely, it corrects $t_1, t_3, t_4, t_5$ couplings.

\vskip 10pt
\noindent
\underline{Diagrams (2h)}

These diagrams are obtained by contracting one vertex (\ref{fig:vertices}o) with the ordinary vertex  (\ref{fig:vertices}a). In all cases 
divergences arise when expanding the propagators at lowest order (self--energy diagrams). They 
are automatically gauge invariant and correct the $t_1, t_3, t_4, t_5$ couplings.

\vskip 10pt
\noindent
\underline{Diagram (2i)}

Finally, possible divergent contributions come from contracting the $\tilde{h}_3$ vertex with
the ordinary $\bar{h}$--vertex in Fig. (\ref{fig:vertices}j). They come from expanding the propagators
up to the first order in $\Gbar$. Even in this case non--trivial cancellations occur and the final
result is the sum of non--vanishing, but gauge invariant contributions to the $t_1, t_3, t_4, t_5$ 
couplings.
 
 \vskip 15pt
The list of diagrams we have analyzed includes all possible divergent diagrams linear in the
deformation parameter. Any other divergence is necessarily proportional to ${\cal F}^2$ and 
comes either from the expansion of the $\ast$--products in the previous diagrams or from new
diagrams constructed from ${\cal F}^2$--vertices in (\ref{main}). Since we know
that any single ${\cal F}^2$ term is automatically gauge--invariant and
appears in the action with its own coupling, we can immediately conclude that the ${\cal F}^2$ 
sector of the action is one--loop renormalizable. 

\vskip 20pt
In conclusion, we have provided evidence that the general action (\ref{main}) is  multiplicatively renormalizable. 
Its renormalization can be then performed by setting
\bea
&& ~~\qquad \qquad \Phi_B^a = Z^{\frac12} \Phi^a \qquad , \qquad \Phib_B^a = \bar{Z}^{\frac12} \Phib^a
\non \\
&& ~~\qquad \qquad \Phi_B^0 = Z^{\frac12} \Phi^0 \qquad , \qquad \Phib_B^0 = \bar{Z}^{\frac12} \Phib^0
\non \\
&&~~\qquad \qquad ~~~~~~~~ (\kappa - 1)_B =Z_{\kappa} (\kappa - 1)
\non \\
&& ~~\qquad \qquad h_B = Z_h h \qquad ~,~ \qquad \bar{h}_B = Z_{\bar{h}} \bar{h}
\non \\
&&~~\qquad \qquad ~~~~~~~~~~~~~~\tilde{h}_{3 \, B} = Z_{\tilde{h}_3} \tilde{h}_3
\non \\
&& ~h_{3\, B}^{(j)} = Z_{h_3^{(j)}} h_3^{(j)} \quad , \quad 
h_{4\, B}^{(j)} = Z_{h_4^{(j)}} h_4^{(j)} \quad , \quad  h_{5\, B}^{(j)} = Z_{h_5^{(j)}} h_5^{(j)} 
\non \\
&& ~~\qquad \qquad t_{n \, B} = Z_{t_n} t_n   \qquad n = 1, \dots, 5
\non \\
&&  ~~\qquad \qquad ~~~~~~~~~~~~~~\tilde{h}_{6 \, B}^{(j)} = Z_{\tilde{h}_6^{(j)}} \tilde{h}_6^{(j)}
\non \\
&&~~ \qquad \qquad  l_{n \, B} = Z_{l_n} l_n   \qquad n = 1, \dots, 12
\eea 
where we have assigned the same renormalization function to the abelian and non--abelian scalar
superfields. 

We consider for instance the nontrivial renormalization of the quadratic matter action, first two lines of 
eq. (\ref{mainmatter}). At one--loop, in terms of renormalized superfields, we can write
\bea
&&\Gamma_{1loop} \rightarrow \intsup  \Big\{  \left( (Z \bar{Z})^{\frac12} - 1 + \frac{a}{\epsilon} \right)
\Tr \left( \Phib \ast \Phi \right)+ 
\\
&& \left( (Z\bar{Z})^{\frac12} Z_{\kappa} - 1 + \frac{b}{\epsilon} \right) \frac{\kappa - 1}{\Nc} 
\Big[ \Tr \Phib  \ast \Tr \Phi 
+ 2i \thb^2 \Fab \Tr (\Gbar_\a^{~\ad} \ast \Phib) \ast \Tr ( \partial_{\beta\ad} \Phi )
\non \\
&~& \qquad \qquad \qquad \qquad \qquad \qquad \qquad \qquad 
+ 2i \thb^2 \Fab \Tr ( \Gbar_\alpha^{~\ad} \ast \Phi) \ast \Tr ( \partial_{\beta\ad} \Phib ) \Big]
\Big\} 
\non
\eea
where, from eqs. (\ref{diver1}, \ref{2pointmodified}) we read 
\bea
&& a = \frac{1}{(4\pi)^2} \, \left[ -2g^2 \mathN + 9 h \bar{h} \left( \mathN + 4 \frac{1-\kappa}{\kappa \mathN} 
\right) \right] 
\non \\
&& b = \frac{1}{(4\pi)^2} \, \frac{1}{\kappa-1} \, \left[ 2 g^2 \mathN + 9 h \bar{h} \left( 1 + 2 \left( 
\frac{1 - \kappa}{\kappa \mathN} \right)^2 \right) \right] 
\eea
In order to cancel divergences we can set 
\bea
&& Z = \bar{Z} =  1 - \frac{1}{(4\pi)^2} \, \frac{1}{\e} \, \left[ -2g^2 \mathN + 9 h \bar{h} \left( \mathN + 4 \frac{1-\kappa}{\kappa \mathN} 
\right) \right] 
\\
&& Z_{\kappa} = 1 + \frac{1}{(4\pi)^2} \, \frac{1}{\e} \, \left[ -2 g^2 \mathN \frac{\kappa}{\kappa -1} 
+ 9 h \bar{h} \mathN  \left( \frac{\kappa-2}{\kappa-1} \right) - 18 \frac{h\bar{h}}{\kappa^2 \mathN}
(2\kappa^2 - \kappa -1) \right] 
\non
\eea
Different choices with $Z \neq \bar{Z}$ are also allowed.

Renormalization of the rest of the couplings then follows, accordingly.

\section{Conclusions}

In this paper we have studied the problem of the renormalizability for nonanticommutative $N=1/2$ 
SYM theories in the presence of interacting matter. The introduction of a superpotential for 
(anti)chiral superfields complicates the investigation of the quantum properties of the gauge theory, 
not only from a technical point of view. In fact, 
at a first sight the non--trivial interplay between partial breaking of supersymmetry, gauge 
invariance of the action and renormalization procedure leads to drastic consequences for the theory: 
In NAC geometry only $SU({\cal N}) \otimes U(1)$ gauge theories are well defined and, as in the 
ordinary case, the renormalization of the kinetic term requires a different renormalization function 
for the $SU({\cal N})$ and $ U(1)$ wave--functions. Consequently, superpotential terms proportional
to the abelian fields need appear with different coupling constants. In superspace formalism 
this can be realized by generalizing the single--trace (anti)chiral interaction to contain different 
trace structures, each one with its own coupling. However, the addition of 
multi--trace terms, while completely harmless in the ordinary SYM theories, in the NAC case
affects the theory in a non--trivial way. 
In fact, these terms are no longer gauge singlets and require 
suitable completions which break explicitly the residual $N=1/2$ supersymmetry.

The way--out we have proposed amounts to re--establish perfect equivalence between 
$SU({\cal N})$ and $U(1)$ wave--function renormalizations by multiplying the abelian 
quadratic term by an extra coupling constant. As a 
consequence, a single--trace superpotential is allowed which respects $N=1/2$ supersymmetry and 
supergauge invariance. 
Basically, we have shifted the problem of deforming the action from the superpotential to the 
K\v{a}hler potential or, in other words, from an integral on chiral variables to an integral on 
the whole superspace. This has the nice effect to leave the residual $N=1/2$ supersymmetry unbroken. 
It is important to stress that in contradistinction with the ordinary case where rescaling the 
abelian kinetic term or suitably rescaling the superpotential couplings lead to equivalent theories, 
in the NAC case this is no longer true. 
In one case we obtain a consistent $N=1/2$ theory whereas in the other case we loose completely 
supersymmetry. The ultimate cause is the non--trivial NAC gauge transformations undergone by the 
abelian superfields. 

Having solved the main problem of adding a matter cubic superpotential we have studied the most general 
divergent structures which could arise at loop level selecting them on the basis of dimensional 
considerations and global symmetries. We have then proposed  the action (\ref{main}) as the most general  
renormalizable gauge--invariant $N=1/2$ deformation of the ordinary SYM field theory with interacting 
matter. 
The next steps should be the complete study of one--loop renormalization,  the computation of 
the $\beta$--functions and the implementation of the massive case. 
Moreover, strictly speaking our results hold only at one--loop. Higher loop calculations 
would be necessary to further confirm the good renormalization properties of our action. 

Generalizing in an obvious way our construction to include more than one (anti)chiral superfields 
would lead to
a consistent NAC generalization of the $N=4$ SYM. This  would be an important step towards clarifying 
the stringy origin of NAC deformations and deformations of the AdS/CFT correspondence.
In particular, it would be nice to investigate how robust properties of $N=4$ SYM like finiteness and 
integrability might be affected by NAC deformations. 

Finally, our approach could be easily applied to the abelian three--field Wess--Zumino  model
studied in \cite{JJP2}.

\section*{Acknowledgements}
\noindent 

This work has been supported in part by INFN, PRIN prot.20075ATT78-002 and the European 
Commission RTN program MRTN--CT--2004--005104. The work of A.R. is supported by the European 
Commission Marie Curie Intra-European Fellowships under the contract N 041443.

\newpage
\appendix

\section{Feynman rules for the general action (\ref{main})} 

In this Appendix we apply the NAC background field method to the action (\ref{main}) and derive
the Feynman rules necessary for calculations of Section 6.  

\vskip 10pt
\noindent
\underline{Gauge sector}

We first concentrate on the gauge sector. As discussed in details in Ref. \cite{GPR2} and reviewed
in Section 2, with the convenient choice of the gauge--fixing action (\ref{gf}),  
in Feynman gauge the covariant gauge propagators are 
\bea
\label{vectorprop}
\langle V^a V^b \rangle &=& g^2\left( \frac{1}{\hat{\Box}} \right)^{ab}
\non \\
\langle V^0 V^0 \rangle &=&  g^2 \left\{\frac{1}{\Tilde{\Box}} \left[ 1 + \left( \frac{g^2}{g^2 + g_0^2} 
\right) \overline{\boldnabla}^\ad \ast \boldnabla^2 \ast \overline{\boldnabla}_\ad \ast
\frac{1}{\Tilde{\Box}} \right] \right\}^{00}
\eea
where $\hat{\Box}, \Tilde{\Box}$ have been defined in (\ref{boxhat}, \ref{tildebox}) 
in terms of $\Box_{cov}$. 
On a generic superfield in the adjoint representation of $SU(\Nc) \otimes U(1)$ we have  
\bea
\label{boxcov2}
(\Box_{cov} \ast \phi)^A &=& \left( \frac{1}{2} \Delb^{\a\ad} \ast \Delb_{\a\ad} \ast
\phi \right)^A 
\non \\
&=& \left( \Box \phi -i [ \Gbar^{\a\ad},  \partial_{\a\ad} \phi ]_\ast - \frac{i}{2}
  [(\partial^{\a\ad} \Gbar_{\a\ad}), \phi]_\ast - \frac{1}{2} [\Gbar^{\a\ad}, [\Gbar_{\a\ad},
  \phi]_\ast ]_\ast \right)^A 
\non  \\
&\equiv& \Box_{cov}^{AB} \ast \phi^B
\eea
Using the general NAC rule
\beq
[F,G]_\ast^A = \frac12 if^{ABC} \{ F^B, G^C \}_\ast + \frac12 d^{ABC} [ F^B, G^C ]_\ast 
\eeq
valid for any couple of field functions in the adjoint representation of the gauge group,
and expanding the $\ast$--product we find
\bea 
\label{boxcov3}
\Box_{cov}^{AB} &=& \Box ~\d^{AB}  + f^{ACB} \Gbar^{C \, \a\ad} \pa_{\a\ad} + 
i d^{ACB} \Fab (\pa_\a \Gbar^{C \, \g\gd}) \pa_\b \pa_{\g\gd}    
-\frac12 f^{ACB} {\cal F}^2 (\pa^2 \Gbar^{C \, \a\ad}) \pa^2 \pa_{\a\ad} 
\non \\
&~&~+~ \cdots 
\eea 
Only the first two terms in (\ref{boxcov2}) have been explicitly indicated. The rest can 
be treated in a similar manner. 

The $\frac{1}{\hat{\Box}}$ and $\frac{1}{\Tilde{\Box}}$
propagators can be expanded in powers of the background fields. We formally write
\bea
&& \frac{1}{\hat{\Box}} = \frac{1}{\Box_{cov}} + \frac{1}{\Box_{cov}} \ast \left( i \tilde{W}^\a 
\Del_\a + i \Wbar^\ad \ast \Delb_\ad \right) \ast \frac{1}{\hat{\Box}}
\non \\
&& \frac{1}{\Tilde{\Box}} = \frac{1}{\Box_{cov}} + \frac{1}{\Box_{cov}} \ast \left( i \tilde{W}^\a 
\Del_\a - \frac{i}{2} (\Delb^\ad \ast \Wbar_\ad) \right) \ast \frac{1}{\Tilde{\Box}}
\eea
Expanding the right hand side we obtain terms proportional to $\tilde{W}_\a, \Wbar_\ad$ 
and terms proportional to the bosonic connections coming from $1/\Box_{cov}$. As follows 
from dimensional considerations and confirmed by 
direct inspection, terms proportional to the field strengths never enter divergent 
diagrams as long as we focus on contributions linear in the NAC parameter. 
Therefore, at this stage we can neglect them. Using the expansion (\ref{boxcov3}) we then find 
\bea
\label{eqn:box_{vector}}
&&\left( \frac{1}{\hat{\Box}} \right)^{ab} , \, \left( \frac{1}{\Tilde{\Box}} \right)^{00}
\rightarrow \left( \frac{1}{\Box_{cov}} \right)^{AB}
\\
\non \\
&&\simeq   \frac{1}{\Box} ~\delta^{AB}
  - \frac{1}{\Box} f^{ACB}~ \Gbar^{C \, \a\ad}~\partial_{\a\ad} \frac{1}{\Box}
- \frac12  \frac{1}{\Box} \, f^{ACD} f^{DEB} \, \Gbar^{C \, \a\ad} \Gbar^E_{\, \a\ad}  \, \frac{1}{\Box}
\non \\  
&~&~~ -  \frac{1}{\Box} id^{ACB}~ \Fab (\pa_\a \Gbar^{C \, \g\gd}) ~ \pa_\b \pa_{\g\gd} \frac{1}{\Box} 
+ \frac12 ~\frac{1}{\Box} ~ f^{ACB} {\cal F}^2 (\pa^2 \Gbar^{C \, \a\ad})~\pa^2 \pa_{\a\ad} 
~\frac{1}{\Box} ~+~ \cdots 
\non  
\eea
In this expression we recognize the ordinary bare propagator $1/\Box$ plus a number of gauge
interaction vertices. 

Further interactions come from the expansion of the remaining terms in (\ref{min1}) or 
(\ref{min2}). Their explicit expression can be found in Appendix E of \cite{GPR2}.  

\vskip 10pt
\noindent
\underline{Matter sector}

We now derive propagators and interaction vertices for the action $S_{matter} + S_{\Gbar} +
S_{\Wbar}$ in (\ref{main}). Since in this paper we are primarily interested in computing divergent 
contributions linear in the NAC parameter, we restrict our analysis to Feynman rules 
which contribute to this kind of terms. In particular, we do not take into account vertices 
proportional to ${\cal F}^2$. 

We first concentrate on the calculation of the chiral propagators.  
As given in eq. (\ref{mainmatter}) the full covariant scalar quadratic term is 
\beq 
\intsup \left\{ \Tr \left( \Phib  \Phi \right) +
\frac{\kappa - 1}{\mathN} \Tr \Phib \Tr \Phi \right\} 
\eeq 
which can be expanded in terms of the background covariantly (anti)chiral fields
(\ref{covchiral}) as
\bea
&& \intsup \left\{ \Tr (\Phibbold \ast e^V \ast \Phibold \ast
  e^{-V}) + \frac{\kappa-1}{\mathN} \Tr(\Phibbold) \Tr(e^V \ast
  \Phibold \ast e^{-V}) \right\} 
  \nonumber \\ 
  &~&= \intsup \left\{ \Tr \left(
  \Phibbold \Phibold + \Phibbold [V, \Phibold]_\ast + \frac{1}{2}
  \Phibbold [V, [V,\Phibold]_\ast]_\ast + \ldots \right) \right. 
  \nonumber \\ 
  &~&~~~~\qquad \qquad \left. + \frac{\kappa-1}{\mathN}  \Tr(\Phibbold)
  \Tr\left(\Phibold + [V, \Phibold]_\ast + \frac{1}{2} [V,
    [V,\Phibold]_\ast]_\ast + \ldots \right) \right\}
\label{exp2}
\eea

We perform the quantum-background splitting 
\beq 
\label{splitting}
\Phibold \rightarrow \Phibold + \Phibold_q, \Phibbold \rightarrow \Phibbold +
\Phibbold_q
\eeq
and concentrate on the evaluation of the quadratic functional integral
\beq
\int \mathcal{D}\Phibold_q \mathcal{D}\Phibbold_q ~ e^{\intsup 
\left\{ \Tr \left( \Phib_q  \Phi_q \right) +
\frac{\kappa - 1}{\mathN} \Tr \Phib_q \Tr \Phi_q \right\} } 
\label{functional}
\eeq

In order to deal with a simpler integral we make the change of variables
\beq
\Phibold_q^{A} \rightarrow \Phibold_q^{'A}=(\Phibold_q^a, \kappa_1
\Phibold_q^0) \qquad , \qquad 
\Phibbold_q^{A} \rightarrow \Phibbold_q^{'A}=(\Phibbold_q^a, \kappa_2 \Phibbold_q^0)
\label{rescaling2}
\eeq
where $\kappa_1$ and $\kappa_2$ are two arbitrary constants satisfying
$\kappa_1 \kappa_2 = \kappa$. The functional integral (\ref{functional}) then takes 
the standard form 
\beq
\int \mathcal{D}\Phibold_q' \mathcal{D}\Phibbold_q' ~ e^{\intsup \Tr \Phibbold_q' \Phibold_q'}
\eeq
We stress that the redefinition (\ref{rescaling2}) in terms of two independent
couplings is admissible because we are working in Euclidean space
where chiral and antichiral fields are not related by complex conjugation.

Adding source terms 
\bea
&& \Tr \intch j\Phibold_q' + \Tr \intach
  \Phibbold_q' \overline j 
\\ 
&~& \qquad \quad = \Tr \intsup \left( j \ast \frac{1}{\Box_+} \ast \boldnabla^2
  \Phibold_q' + \Phibbold_q' \ast \frac{1}{\Box_-} \ast \overline{\boldnabla}^2 \ast \overline j
  \right)
  \non
\eea  
with $\Box_\pm$ defined in (\ref{boxes}), and taking into account the complete action
the quantum partition function reads 
\bea
\mathbf Z(j, \overline j) &=& e^{S_{int}(\frac{\delta}{\delta j},
    \frac{\delta}{\delta \overline j})} \int \mathcal{D}\Phibold_q'
  \mathcal{D}\Phibbold_q'  \exp{ \Tr \intsup \left[
      \Phibbold_q' \Phibold_q' \right.}
\\      
&~& \qquad \qquad   \qquad \qquad  \qquad \qquad  \qquad \qquad {\left.  + j  \ast \frac{1}{\Box_+} \ast \boldnabla^2
      \Phibold_q' + \Phibbold_q' \ast \frac{1}{\Box_-} \ast \overline{\boldnabla}^2
      \ast \overline j \right]}
\non 
\label{partition}
\eea
Here $S_{int}$ contains all gauge--scalar fields interaction vertices in (\ref{exp2}) plus
interactions coming from the rest of terms in $S_{matter} + S_{\Gbar} + S_{\Wbar}$. 

We can perform the Gaussian integral in (\ref{partition}) by standard techniques, obtaining the
NAC generalization of the usual superspace expression \cite{superspace}
\beq
 \label{eqn:part_func}
  \mathbf Z = \Delta \ast  e^{S_{int}( \frac{\delta}{\delta j},
    \frac{\delta}{\delta \overline j})} \exp{\left( - \intsup j \ast
    \frac{1}{\Box_-} \ast \overline j \right)}
\eeq
where $\Delta$ is the functional determinant
\beq
\label{eqn:Delta}
  \Delta = \int {\cal D} \Phibold_q' {\cal D} \Phibbold_q' ~ \exp{\Tr
    \intsup \Phibbold_q' \Phibold_q'}
\eeq
which contributes to the gauge effective action \cite{GPR2}. 

From the expression (\ref{eqn:part_func}) we can read the covariant propagators for prime
superfields
\beq
\langle \Phibold_q'^A \Phibbold_q'^B \rangle = - \left( \frac{1}{\Box_-} \right)^{AB} 
\eeq
which, in terms of the original $\Phi, \Phib$ superfields gives
\bea
\label{rescaledprop1}
  \langle \Phibold_q^a \Phibbold_q^b \rangle &=& - \left(
  \frac{1}{\Box_-} \right)^{ab} 
\\ 
 \label{rescaledprop2} 
  \langle \Phibold^0
  \Phibbold^b \rangle &=& - \frac{1}{\kappa_2} \left(
  \frac{1}{\Box_-} \right)^{0b} 
\\ 
  \label{rescaledprop3}
  \langle \Phibold^a
  \Phibbold^0 \rangle &=& - \frac{1}{\kappa_1} \left(
  \frac{1}{\Box_-} \right)^{a0} 
\\ 
  \label{rescaledprop4}
  \langle \Phibold^0
  \Phibbold^0 \rangle &=& - \frac{1}{\kappa} \left(
  \frac{1}{\Box_-} \right)^{00}
\eea

The expansion of the scalar covariant propagators can be performed following a prescription
similar to the one used for the gauge propagator. We can formally write
\beq
\frac{1}{\Box_-} = \frac{1}{\Box_{cov}} + \frac{1}{\Box_{cov}} \ast \left(i \Wbar^\ad \ast \Delb_\ad 
+ \frac{i}{2} (\Delb^\ad \ast \Wbar_\ad) \right) \ast \frac{1}{\Box_-}
\eeq
Since terms proportional to the field strengths never enter divergent diagrams linear in $\Fab$, we can 
neglect them and write
\bea
\label{eqn:box_{cov}}
&&\left( \frac{1}{\Box_-} \right)^{AB} \rightarrow \left( \frac{1}{\Box_{cov}} \right)^{AB}
\\
\non \\
&&\simeq   \frac{1}{\Box} ~\delta^{AB}
  - \frac{1}{\Box} f^{ACB}~ \Gbar^{C \, \a\ad}~\partial_{\a\ad} \frac{1}{\Box}
- \frac12  \frac{1}{\Box} \, f^{ACD} f^{DEB} \, \Gbar^{C \, \a\ad} \Gbar^E_{\, \a\ad}  \, \frac{1}{\Box}
\non \\  
&~&~~ -  \frac{1}{\Box} id^{ACB}~ \Fab (\pa_\a \Gbar^{C \, \g\gd}) ~ \pa_\b \pa_{\g\gd} \frac{1}{\Box} 
+ \frac12 ~\frac{1}{\Box} ~ f^{ACB} {\cal F}^2 (\pa^2 \Gbar^{C \, \a\ad})~\pa^2 \pa_{\a\ad} 
~\frac{1}{\Box} ~+~ \cdots 
\non  
\eea
The first term is diagonal in the color indices and gives the ordinary bare propagator. The rest 
provides interaction vertices between scalars and gauge superfields. 

From the expansion (\ref{eqn:box_{cov}}) it is clear that the mixed propagators 
(\ref{rescaledprop2}, \ref{rescaledprop3}) are always 
proportional to the NAC parameter, according to the fact that in the $N=1$ limit they need vanish. 
It follows that the dependence on the $\kappa_1$ and $\kappa_2$ couplings is peculiar of the 
NAC theory, whereas in the ordinary limit only their product $\kappa$ survives.

Additional interaction terms are contained in $S_{int}$ and arise from the background field
expansion of the full action $S_{matter} + S_{\Gbar} + S_{\Wbar}$. We now describe the correct way 
to obtain such vertices concentrating only on the ones at most linear in $\Fab$.

We begin by considering $S_{matter}$. From the quadratic action $\intsup \Tr \Phib \Phi$, 
after the expansion 
(\ref{exp2}) and the shift (\ref{splitting}) we obtain (3a,3b)--type vertices in Fig. 3 where $V$ is
quantum and $\Phibold$ and/or $\Phibbold$ are background. 
Expanding the $\ast$--products ordinary vertices plus vertices proportional to $\Fab$ and 
${\cal F}^2$ arise.

We then consider the $(\kappa -1)$ terms in (\ref{mainmatter}) 
\bea
\label{kappa1}
&&\frac{\kappa - 1}{\Nc}\intsup \Big[ \Tr \Phib  \ast \Tr \Phi
\\
&~~&~~~~~~~~~~~~\quad \quad + 2i \thb^2 \Fab 
\Tr (\Gbar_\a^{~\ad} \ast \Phib) \ast \Tr ( \partial_{\beta\ad} \Phi )
+ 2i \thb^2 \Fab \Tr ( \Gbar_\alpha^{~\ad} \ast \Phi) \ast \Tr ( \partial_{\beta\ad} \Phib ) \Big]
\non
\eea 
We expand the (anti)chiral superfields as
\beq
\label{shift}
\Phi \rightarrow \Phibold + \Phibold_q + [ V, \Phibold + \Phibold_q]_\ast + \frac12 [V, [V, 
\Phibold + \Phibold_q]_\ast]_\ast \quad , \quad \Phib \rightarrow \Phibbold + \Phibbold_q
\eeq
and, at the same order in $V$, the gauge connection as 
\beq
\label{gammasplitting}
  \Gbar_{\a\ad} \rightarrow \Gbar_{\a\ad}
  - \Del_\a \left[ \overline{\boldnabla}_{\ad} , V \right]_\ast
  + \frac{1}{2} \Del_\a \left[ \left[\overline{\boldnabla}_{\ad} ,
  V \right]_{\ast} , V \right]_{\ast}
\eeq
Collecting the various terms we generate (3c,3d)--vertices 
in Fig. 3 with background gauge connections and quantum matter plus (3e,3f,3g)--vertices with 
quantum gauge and $\Phibold$ or $\Phibbold$ background. 

As a nontrivial example, we derive in details the contributions (3e,3f,3g). 
Forgetting for a while the superspace integration and the 
overall coupling constant and writing $\pa_\a = \Del_\a - i \thb^\ad \pa_{\a\ad}$, 
from the first term in (\ref{kappa1}) we have
\bea
\label{first}
 && \Tr ([V, \Phibold]_\ast) \, \Tr \Phibbold \rightarrow -\Fab \, \Tr
  ([\partial_\a V, \partial_\b \Phibold]) \, \Tr \Phibbold 
  \non \\ 
&&
\non \\  
  &~& \rightarrow -2i \Fab \, \thb^\ad \, \Tr ( V \Dela \Phibold) \, \Tr \partial_{\b\ad} \Phibbold 
- 2\Fab \thb^2 \, \Tr ( \partial_\a^{~\ad} V \, \Phibold) \, \Tr \partial_{\b\ad} \Phibbold
\eea
where superspace total derivatives have been neglected and $\Phibold, \Phibbold$ stand for either
quantum or background. 

Using the expansion (\ref{gammasplitting}) the second term in (\ref{kappa1}) gives a contribution of the form
\beq 
\label{eqn:(k-1)2}
  2i \Fab \, \thb^2 \Tr ( \Gbar_\a^{~\ad} \Phibbold) \, \Tr ( \partial_{\beta\ad} \Phibold) 
  \rightarrow  - 2i \Fab \, \thb^2 \Tr ( [\Delb^\ad,
  V] \, \Phibbold) \, \Tr ( \partial_{\beta\ad} \, \Dela \Phibold )
\eeq
Similarly, the third term in (\ref{kappa1}) gives  
\bea
\label{third}
&& 2i \Fab \thb^2 \Tr ( \Gbar_\a^{~\ad} \Phibold) \, \Tr ( \partial_{\beta\ad} \Phibbold
  ) \rightarrow 2i \Fab \thb^2  \left\{ \Tr ( \Gbar_\a^{~\ad} [V, \Phibold]) \, \Tr
  ( \partial_{\beta\ad} \Phibbold ) \right.
\non \\  
&& \left. - \Tr ( \Dela \overline{D}^\ad
  V \, \Phibold) \, \Tr ( \partial_{\beta\ad} \Phibbold ) + \Tr (
  [V, \Gbar_\a^{~\ad}] \Phibold) \, \Tr ( \partial_{\beta\ad} \Phibbold ) -
  i \Tr ( [\Gbar^\ad, \Dela V] \Phibold) \, \Tr
  ( \partial_{\beta\ad} \Phibbold ) \right\}
  \non \\ 
&& \non \\
&& ~~~~ = 2 \Fab \thb^2 \Tr( \partial_\a^{~\ad} V \, \Phibold) \, \Tr ( \partial_{\beta\ad} \Phibbold) 
 + 2i \Fab \, \thb^2 \, \Tr ( [\Delb^\ad, \Del_\a V] \, \Phibold) \, \Tr
  ( \partial_{\beta\ad} \Phibbold )
\eea
Summing the three contributions a nontrivial cancellation occurs between the second term in
(\ref{first}) and the first term in (\ref{third}) and we are left with
\bea
\frac{\kappa-1}{\Nc} &\int& d^4x d^4 \th\left\{-2i \Fab \, \thb^\ad \, \Tr ( V \Dela \Phibold) \, 
\Tr \partial_{\b\ad} \Phibbold  
- 2i \Fab \, \thb^2 \Tr ( [\Delb^\ad,
  V] \, \Phibbold) \, \Tr ( \partial_{\beta\ad} \, \Dela \Phibold ) \right. 
\non \\
&~&~~~~~~~~~~~+ \left. 2i \Fab \, \thb^2 \, \Tr ( [\Delb^\ad, \Del_\a V] \, \Phibold) \, \Tr
  ( \partial_{\beta\ad} \Phibbold ) \right\}
\eea
which correspond to the three vertices (3e, 3f, 3g). 

The rest of terms in the $S_{matter}$ can be easily treated by the shift (\ref{shift}). Neglecting
${\cal F}^2$ contributions only the superpotential and the $\tilde{h}_3$ term survive and 
lead to pure matter vertices of the form (3h, 3i, 3j, 3k, 3l) and the mixed vertex (3m). 

We now turn to $S_{\Gbar}$ and briefly sketch the quantization of $t_j$ vertices. 
At linear order in the NAC parameter we can forget the $\ast$--product
in the commutators of $t_3, t_4, t_5$ terms.  
We perform the shift (\ref{shift}) on the (anti)chirals and (\ref{gammasplitting}) on the connection. 
In particular, for the gauge invariant linear combination appearing in $t_3, t_4, t_5$ terms we have
\beq
  \partial_{\beta\ad} \Gbar_\alpha^{~\ad} - \frac{i}{2}
    [ \Gbar_{\beta\ad}, \Gbar_\alpha^{~\ad} ]
\, \longrightarrow \,  
  \partial_{\beta\ad} \Gbar_\alpha^{~\ad} - \frac{i}{2} [
    \Gbar_{\beta\ad},  \Gbar_\alpha^{~\ad} ] -
  \overline{\boldnabla}_{\beta\ad} \boldnabla_\alpha \overline{\boldnabla}^\ad \;  V
\label{eqn:ti quant}
\eeq
Collecting only the contributions which may contribute at one--loop we produce the (3n) vertex
in Fig. 3 where matter is quantum and (3o, 3p) vertices where $\Phi$ or $\Phib$ are quantum.  
We note that they all exhibit a gauge--invariant background dependence.

\begin{figure}
  \includegraphics[width=\textwidth]{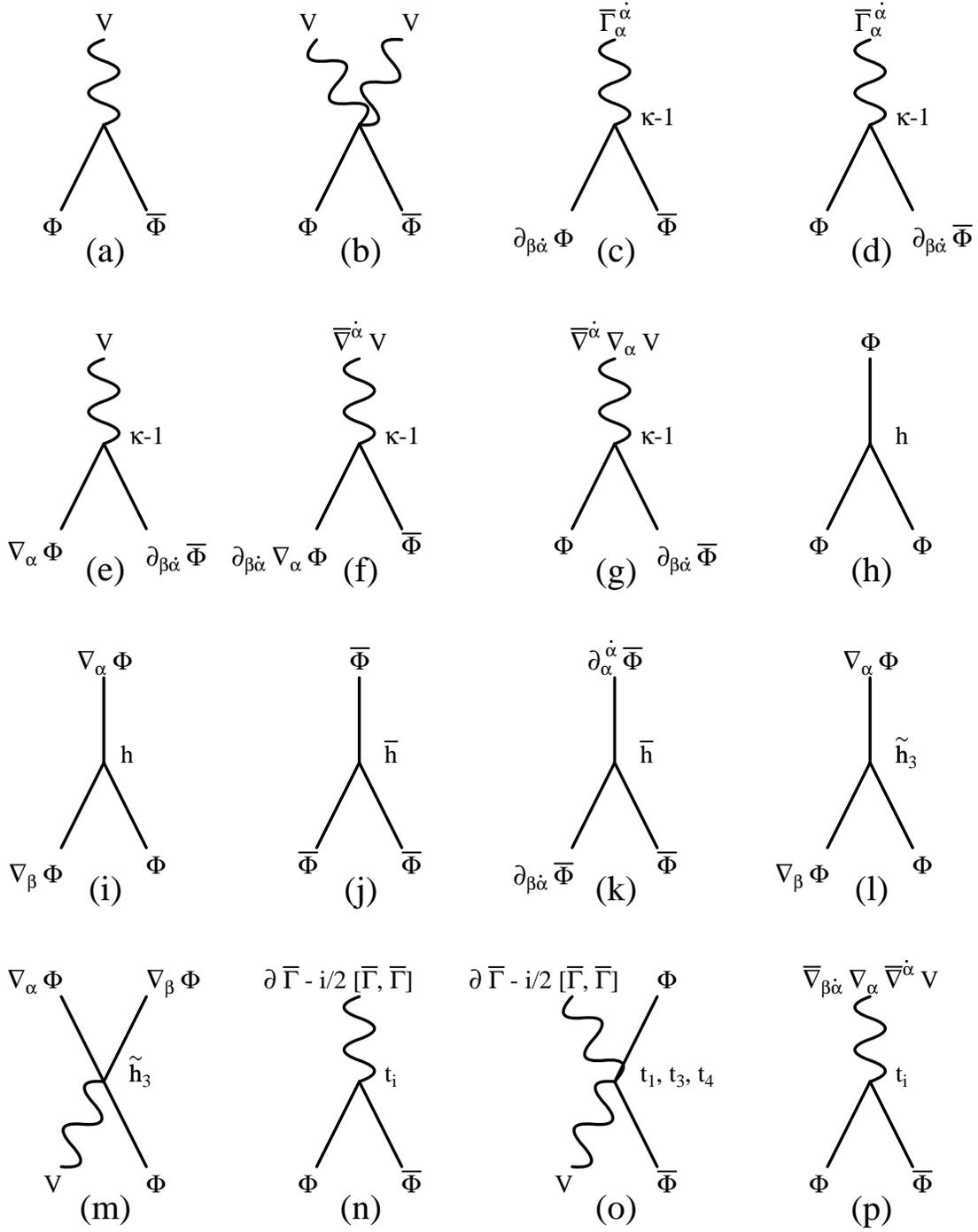} 
  \caption{Vertices from the action (\ref{main}) at most linear in the NAC parameter $\Fab$.  
  The (a,b,h,j)--vertices are order zero in $\thb$, the (e)--vertex is proportional to $\thb^\ad$ whereas 
  the remaining vertices are all proportional to $\thb^2$.} 
  \label{fig:vertices}
\end{figure}

\newpage


\end{document}